\documentclass[fleqn,usenatbib]{mnras}

\usepackage{newtxtext,newtxmath}
\usepackage[T1]{fontenc}

\usepackage{graphicx}	% Including figure files
\usepackage{amsmath}	% Advanced maths commands
\usepackage{gensymb}
\usepackage{comment}
\usepackage{booktabs}
\title[Partial-coherence model]{Pulsar polarization: a partial-coherence model}

\author[L. S. Oswald et al.]{
L. S. Oswald$^{1,2}$\thanks{E-mail: lucy.oswald@physics.ox.ac.uk (LSO)},
A. Karastergiou$^{1}$, 
S. Johnston$^{3}$, 
\\
$^{1}$Department of Astrophysics, University of Oxford, Denys Wilkinson Building, Keble Road, Oxford OX1 3RH, UK\\
$^{2}$Magdalen College, University of Oxford, Oxford OX1 4AU, UK\\
$^{3}$Australia Telescope National Facility, CSIRO, Space and Astronomy, PO Box 76, Epping, NSW 1710, Australia\\
}

\date{Accepted 2023 July 21. Received 2023 July 4; in original form 2023 January 12}

\pubyear{2023}

\begin{document}
\label{firstpage}
\pagerange{\pageref{firstpage}--\pageref{lastpage}}
\maketitle

\begin{abstract}
The population of radio pulsars is observed to demonstrate certain polarization properties not explained by the conventional picture of pulsar polarization, namely frequency evolution of polarization, deviations of the linear polarization angle from a curve of geometric origins and the presence of features in the circular polarization. We present the partial-coherence model as a way to explain the co-occurrence of these features and to provide an origin for circular polarization in radio pulsar profiles. We describe the mathematics of the model and demonstrate how it can explain these observed features, both on a population level and for the idiosyncrasies of individual pulsars. The partial coherence model can account for complex polarization behaviour, enabling improved access to information about pulsar geometries. We discuss the scientific implications of this for our understanding of pulsar radio emission and propagation. 
\end{abstract}


%%%%%%%%%%%%%%%%%%%%%%%%%%%%%%%%%%%%%%%%%%%%%%%%%%

%%%%%%%%%%%%%%%%% BODY OF PAPER %%%%%%%%%%%%%%%%%%

\section{Introduction}
Polarimetric observations have long been a source of important information of the physics of radio pulsars. The angle of linear polarization (position angle or PA) was identified as being related to the geometry of the pulsar beam through the rotating vector model (RVM) by \cite{Radhakrishnan1969}. Subsequent observations of PA profiles revealed jumps of $90\degree$ which are evidence of the pulsar radio emission being composed of two orthogonally polarized modes of emission (OPMs) \citep[e.g.][]{Manchester1975}. \cite{Michel1987} theorised that the handedness of circular polarization should change with viewing direction, something experimentally endorsed by the circular polarization handedness measurements of the precessing pulsar J1906$+$0746 \citep{Desvignes2019b} and observations of pulsars with interpulses \citep{Johnston2019a}. These papers also provide the best observational evidence for the RVM as an explanation for the shape of pulsar PA profiles. Considerable observational evidence for the incoherent combination of orthogonal modes has also been documented since that first evidence of 90$\degree$ PA jumps, including studies of single pulse polarization states \citep[e.g.][]{CordesRankinBacker1978} and the appearance of a new component in the pulse profile of PSR~J0738$-$4042 with polarization properties that are orthogonal to those of the surrounding profile \citep{Karastergiou2011}. However, this description is insufficient to explain the full variety of polarization features seen across the pulsar population, which includes not only orthogonal jumps, but also other complexities in both the PA profile and in the frequency-dependent polarization evolution.

Theoretical understanding of pulsar emission has developed alongside this profileration of observational evidence. We defer a detailed explanation of the theoretical context to Section \ref{sec:history}, but a qualitative description is useful for showing how new observations can be best used to extend that understanding. That description is as follows. 
Two orthogonal modes of emission are generated in the pulsar magnetosphere; these propagate through the magnetosphere, likely along different paths and experiencing refraction with different refractive indices; they may transform into each other coherently, either along the propagation path or in some polarization limiting region; generalized Faraday rotation may be important; and there exists a polarization limiting radius where the plasma density is low enough that there is no longer any appreciable influence of the pulsar's magnetosphere on the polarization of the radio emission. From this point the polarization state may be considered fixed and any further effects are the result of propagation through the interstellar medium. Under this description, the polarization observed at a given pulse phase and frequency will depend on the following: the nature of the paths and refractive effects on the modes and of their interactions with each other, combined with the projection effects of how the observer's viewing angle cuts the beam at a given pulse phase, and finally the nature of observations as being either single pulses or the combination of many single pulses to create an integrated profile. Studies of both individual pulsars and of the pulsar population as a whole provide information about the nature of any such propagation effects. 

In \cite{Oswald2023} we presented an in-depth study of the polarization properties of 271 pulsars as captured by broad-band observations with Murriyang, the Parkes radio telescope. Our particular focus was to categorize polarization behaviour not accounted for in the traditional picture of pulsar radio polarization. We found that circular polarization played a key role in the overall polarization properties of a pulsar, and described how these properties evolved with $\dot{E}$ across the population. In particular, we identified a link between the presence of circular polarization features in pulse profiles and departures of the PA profile from a RVM-like shape, something that was also linked with frequency evolution of polarization. Furthermore, frequency- and phase-evolution of polarization is seen more for pulsars with low $\dot{E}$ (particularly deviations from the RVM), along with low polarization fractions and a large proportion of that polarization being circularly polarized, whereas higher-$\dot{E}$ pulsars have simpler profiles and are strongly linearly polarized, as has been previously noted by \cite{Johnston2006}. We also showed that frequency evolution of pulsar polarization tends to follow similar evolution from low to high frequency as is seen across the pulsar population from high to low $\dot{E}$: pulsars simultaneously depolarize and increase the relative contribution of circular polarization.

One idea which can explain these characteristics is that the orthogonal modes do not simply add incoherently, but also have a coherent addition contribution. This would mean that the observed polarization is the incoherent sum of three contributions: the two modes and their coherent combination.
Recent studies of single pulses for individual pulsars have demonstrated complex polarization effects that are well explained as resulting from coherent mode interaction: see for example the work of \cite{Dyks2019} and \cite{Primak2021}. It is therefore important to consider the polarization behaviour of the whole pulsar population, to investigate whether polarization complexities more generally can be similarly explained in this way.

In this paper we present a simple and straightforwardly motivated way of parametrizing the diverse effects described in \cite{Oswald2023} as resulting from partially coherent orthogonal mode interaction. We define a model which captures this partially-coherent behaviour through just three parameters: the ratio of the orthogonal modes' strengths; the phase offset between the modes; and the extent to which they interact coherently rather than incoherently. We then test whether allowing these parameters to evolve in a sensible way can account for the polarization effects we observe in the pulsar population. 

In Section \ref{sec:model}, we describe the motivations for our parametrization and explain the observational effects that will result from varying each of the three parameters in the model. We describe how the observational results from \cite{Oswald2023} can be interpreted in the context of this model in Section \ref{sec:obs_interpretation}. In Section \ref{sec:indivpsrs}, we show how the mathematics of the model can be applied to convert observed Stokes parameters to mode interaction parameters. Applying this to three individual sources, we demonstrate that their complex polarization behaviour is simply explained as a smooth varying of one of the three model parameters with the other two held constant. Finally in Sections \ref{sec:discussion} and \ref{sec:conc} we discuss the extent to which the model describes the observed polarization behaviour of radio pulsars, consider the physics underlying the model in the context of historical theoretical perspectives and elaborate on directions for future work.

\section{The partial-coherence model}
\label{sec:model}

We define the partial-coherence polarization model: that pulsar polarization at a given wavelength is the result of two orthogonally polarized modes of emission which combine both coherently and incoherently to produce the observed polarization. This means that the observed polarization is made up of the incoherent sum of three contributions: the two orthogonally polarized modes and a third contribution made up of the coherent sum of these two modes. 
This picture is motivated by simplicity: the model relies only on three parameters. These are the mode strength ratio $R$, mode phase offset $\eta$ and coherence fraction $C$. In practice these parameters are always applied to finite bandwidth data: it is therefore necessary to use narrow bandwidths when comparing observations to model parameters. The two orthogonal modes may be either entirely linearly polarized or elliptically polarized: in the latter case there will be some contribution to circular polarization from the initial mode ellipticity in addition to that resulting from the coherent addition. We describe the behaviour of each model parameter in general terms before investigating how observations can be interpreted within this model, both on a population scale and for individual sources. We discuss the relevant mathematics of the model in the contexts of these observations and defer the full derivations to Appendices \ref{app:maths}, \ref{app:appendixgamma} and \ref{app:appendixC}.

\subsection{Mode strength ratio}

When considering polarization as a fraction of total intensity, information about the actual strengths of the two orthogonal modes is unnecessary; however, the ratio of those strengths $R$ will clearly be important. The impact of relative mode strength is clearest when considering the two extreme cases: equal strength modes ($R = 1$) and one mode completely dominating the other. When one mode dominates, the observed polarization will be 100\% polarized.  When the two modes are equal, the observed polarization could be anything from 100\% polarized to completely unpolarized, and could have any level of circular polarization, depending on the coherence fraction and phase offset parameters.

\subsection{Mode phase offset}

The phase offset $\eta$ between the two orthogonal modes dictates the amount of circular polarization generated by their coherent combination. In the case of linearly polarized orthogonal modes, if $\eta = 0\degree$, then coherent addition of the modes will simply lead to linearly polarized radiation rotated by some angle. If $\eta = 90\degree$, then all of the coherently added radiation will be circularly polarized. For elliptical modes equivalent behaviour is seen, but there will be some circular polarization even when $\eta = 0\degree$. 

\subsection{Coherence fraction}

The coherence fraction $C$ defines what proportion of the two modes is added coherently and what proportion gives an incoherent contribution to the total polarization observed. As described in more detail in Appendix \ref{app:maths}, the final polarization from this model consists of three Stokes parameter contributions added together. These are the individual contributions from each orthogonal mode, and the contribution from the coherent combination of the modes. The relative fraction of the coherently combined part, compared to the original modes, is given by $C$. 

As a result, the coherence fraction, combined with the mode strength ratio, dictates the overall polarization fraction of the observed polarization. If $C = 0$, then the polarization fraction will be dependent entirely on the strength ratio of the two incoherently combined modes, $R$: for example, equal strength modes would result in no net polarization. If $C=1$ then the resultant pulse profile will be 100\% polarized, with the angle of linear polarization and the amount of circular polarization being dependent on $R$ and $\eta$.

\section{Interpretation of observations within the partial-coherence model}
\label{sec:obs_interpretation}

In the partial-coherence model picture, the observed polarization state is expected to result from a combination of two effects. First, the pulsar's rotation and its geometry dictate the sweep of the PA profile as described in the rotating vector model (RVM). Then, the partial-coherence model describes the rotation of the PA away from that RVM curve, alongside the generation of circular polarization. The observed link between deviations from the RVM in the PA profile and features in the circular polarization, as described in \cite{Oswald2023}, is arguably the clearest evidence for a coherent addition explanation of the origin of some, or possibly all, of the circular polarization in pulse profiles. Coherently combining two orthogonal modes with $0\degree < \eta < 90\degree$ will result in rotation of the linear polarization angle and simultaneous introduction of circular polarization. If the PA of a mode is set by the RVM, then the coherent combination will lead to rotation that manifests as deviation from that RVM profile at a given phase bin. 

The link between RVM deviations of the PA and the presence of circular polarization was also found to be related to pulse profiles that exhibited frequency evolution of the polarization. If the PA is set by the pulsar geometry alone, then there should be no evolution of the PA with frequency. However, it is easy to include frequency evolution in the partial-coherence model, simply by allowing the mode interaction parameters to evolve with frequency. We now consider more carefully how the parameters of the partial-coherence model map onto observable polarization properties, and interpret the frequency-dependent polarization behaviour of the pulsar population in that context.

\subsection{Mapping model parameters onto polarization observations}
\label{sec:map}
We make use of four key observable parameters defined in \cite{Oswald2023} for comparing polarization behaviour with the parameters of the partial-coherence model. These are the total, linear and absolute circular polarizations as a fraction of intensity ($p$, $l$ and $\lvert v\rvert$) and the circular contribution $\theta$. The pairings $l$--$\vert v\rvert$ and $p$--$\theta$ can be derived from each other and all are derived from the Stokes parameters $I$, $Q$, $U$ and $V$. The definitions are as follows: 
\begin{align}
        l &= \sqrt{Q^{2} + U^{2}}/I & &\equiv p\cos(\theta) 
        \label{eq:l_intro} \\
        \lvert v\rvert &= \lvert V\rvert/I & &\equiv p\sin{(\theta)}
        \label{eq:v_intro} \\
        p &= \sqrt{Q^{2} + U^{2} + V^{2}}/I & &\equiv \sqrt{l^{2} + \lvert v\rvert^{2}} 
        \label{eq:p_intro} \\
        \theta &= \arctan{(\lvert V\rvert/\sqrt{Q^{2} + U^{2}})} & &\equiv \arctan{(\lvert v\rvert/l)}. 
        \label{eq:theta_intro}
\end{align}

Varying the model parameters $R$, $\eta$ and $C$ leads to variations in the linear and circular polarization fractions $l$ and $\lvert v\rvert$ (and hence $p$ and $\theta$), and to the angle of linear polarization (PA) and handedness of circular polarization. Allowing the model parameters to evolve smoothly across both pulse phase and frequency for individual pulsars, and across the population as a whole, would lead to a large range of predicted polarization behaviours. We focus first on the observed behaviour of the population as a whole, before moving onto detailed study of individual pulsars in Section \ref{sec:indivpsrs}.

\begin{figure*}
    \centering
    \includegraphics[width=\textwidth]{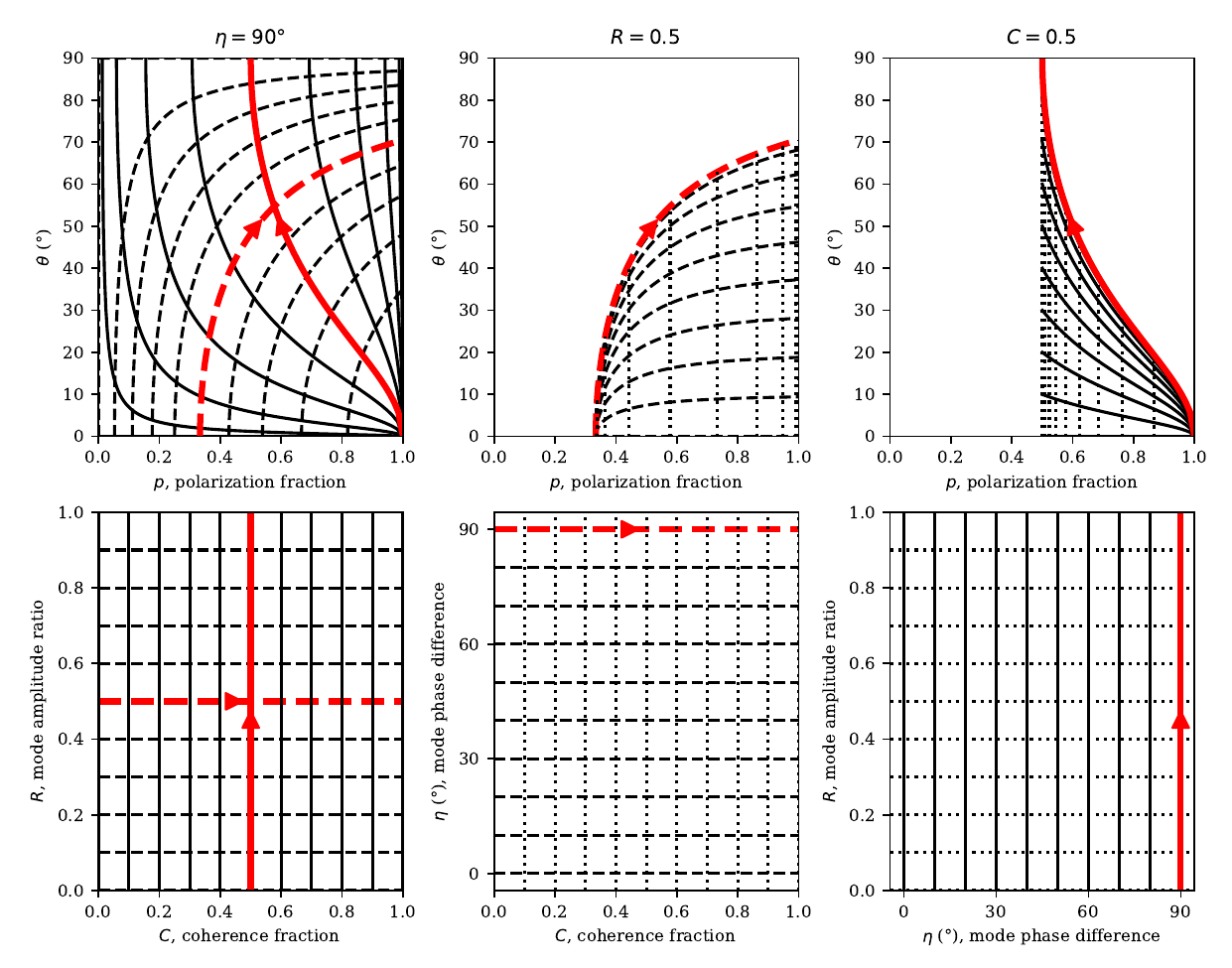}
    \caption{Grids showing the relationship between observable polarization parameters $p$ and $\theta$ and partial-coherence model parameters $R$, $\eta$ and $C$ (all defined in the text). The bottom three plots show grids of pairs of partial-coherence model parameters (the third being held at a constant value indicated in the subplot title), indicated with dotted, dashed and solid lines. The top three plots show how those grids map onto the $p$--$\theta$ diagram. The same line patterns are used throughout: lines of varying $R$ are solid; lines of varying $C$ are dashed and lines of varying $\eta$ are dotted. Thicker red lines are used to indicate a particular set of parameters, with the arrows indicating the direction of increase of a parameter.}
    \label{fig:RetaCptheta}
\end{figure*}

Fig. \ref{fig:RetaCptheta} shows how the $R$, $\eta$ and $C$ parameters map onto the parameter space of the measurable polarization properties $p$ and $\theta$, in the case where the orthogonal modes are entirely linearly polarized. We focus on linearly polarized modes as the simplest case and then discuss how the observational picture is modified for elliptical modes. In each subplot, the solid lines show varying $R$ with $C$ and $\eta$ held constant; the dashed lines show varying $C$ with $R$ and $\eta$ held consant, and the dotted lines show varying $\eta$ with $R$ and $C$ held constant. The arrows mark the directions of the variables increasing. We can see from the curved gridlines in the top-left subplot that small values of $R$ (close to $0$) correspond to high polarization fraction ($p$ close to 1) that is strongly linear ($\theta$ close to 0). Large values of $R$ (close to 1) and $C$ are required to attain strongly circularly polarized polarization ($\theta$ close to 90$\degree$).

Whereas varying $R$ and $C$ causes the observed polarization parameters to follow curved tracks in the $p$--$\theta$ diagram, varying $\eta$ (with $R$ and $C$ constant) has no impact on polarization fraction $p$. This leads to a vertical track on the $p$--$\theta$ diagram (see dotted lines in top-middle and top-right subplots). We can search for ordered behaviour in the $p$--$\theta$ diagram, such as vertical lines of this nature, and infer its link to these model parameters. For example, horizontal tracks with $\theta \lesssim 15\degree$ would result from varying $R$ (or possibly varying $C$, but only if $\eta$ is very small). Tracks following curved paths from bottom right towards top left would also indicate varying $R$, whereas tracks following curved paths from bottom left to top right would indicate varying $C$. Vertical tracks would indicate varying $\eta$, unless $p \lesssim 0.3$ or $p \gtrsim 0.8$, in which case it could also be varying $R$. 

\begin{figure*}
    \centering
    \includegraphics[width=\textwidth]{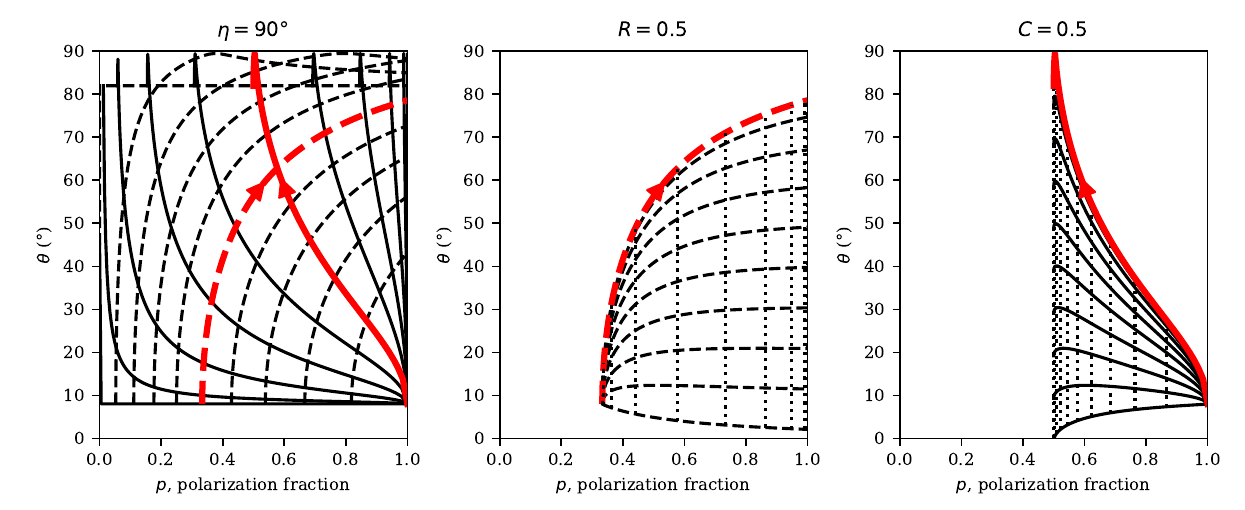}
    \caption{Figure set up as described in Fig. \ref{fig:RetaCptheta}, but now showing the impact on observed polarization of using elliptically polarized natural modes. The ellipticity set up for this example is described with parameters $\alpha = -\beta = 0.1$ and $\chi = -\psi = 45\degree$ (see Appendix \ref{app:sec:ellip} for an explanation of these parameters).} 
    \label{fig:grids_with_ellipticity}
\end{figure*}

Fig. \ref{fig:grids_with_ellipticity} shows how the $R$--$\eta$--$C$ gridlines are altered in the cases of elliptical orthogonal modes with increasing degrees of ellipticity. Essentially, the gridlines are shifted vertically to account for the fact that circular polarization is present even in the absence of coherent addition, but the behaviour is otherwise unchanged. This justifies our approach of keeping the model conceptually simple by focusing on linearly-polarized orthogonal modes in our comparisons with observations in the coming sections. 

\subsection{Frequency evolution of polarization: comparing past interpretations and modern observations with the partial-coherence model}
We observed in \cite{Oswald2023} that pulsars tend to depolarize with increasing frequency ($p$ decreases) whilst circular contribution $\theta$ increases. \cite{Karastergiou2005} proposed that the frequency-dependence of orthogonal jumps in pulsar profiles, and the depolarization at increasing frequency, could be explained by the two (incoherently-combining) modes having different spectral indices. This would lead directly to an evolution of $R$ with frequency in our description. \cite{Gangadhara1997} suggested that, as increasing frequency reduces the coherence-length for mode interaction, modes will combine together more incoherently at higher frequencies: this we can conceptualize as a decrease in $C$ with frequency. 

The explanations given in those papers do not directly address the relationship between depolarization and increasing relative contribution of circular polarization. However, in the partial-coherence model, the two are directly linked to each other. 
Fixing $C$ and $\eta$ and allowing $R$ to vary causes $p$ and $\theta$ to track along a curved line that tends to go from high-$p$, low-$\theta$ to low-$p$, high-$\theta$. An example of this is indicated by the solid thicker red line on the top-left subplot of Fig. \ref{fig:RetaCptheta}, which shows varying $R$ for $C = 0.5$ and $\eta = 90\degree$. The same solid red line is also indicated on the top-right, bottom-left and bottom-right subplots as part of the full mapping of this set of 3 parameters.

\begin{figure}
    \centering
    \includegraphics[width=\columnwidth]{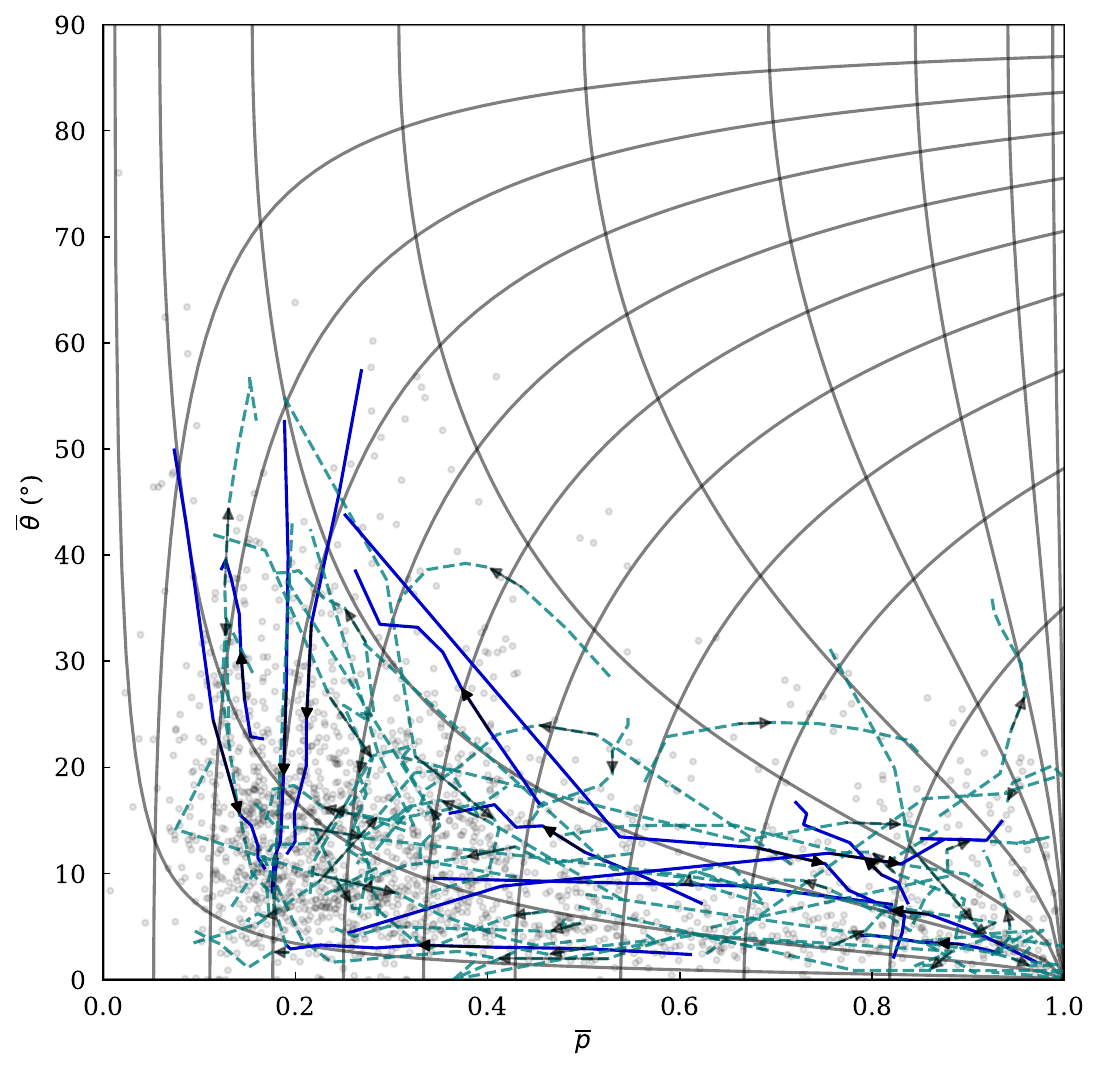}
    \caption{A reproduction of fig. 7 from Oswald et al (2023) showing the frequency-evolution of polarization on the $p$--$\theta$ diagram for the observed pulsar sample of that paper. Overlaid is the grid of lines of varying $R$ and $C$ for $\eta = 90\degree$, as visualized in the top-left subplot of Fig. \ref{fig:RetaCptheta}. }
    \label{fig:RetaCpthetaDATA}
\end{figure}

In Fig. \ref{fig:RetaCpthetaDATA} we show the frequency-dependent data results taken from fig. 7 of \cite{Oswald2023} plotted on a graph of $\bar{\theta}$ vs. $\bar{p}$ (both parameters averaged across pulse phase), but now with the $R$--$C$ parameter-space grid (with $\eta = 90\degree$) mapped on top of it. It can be seen that, qualitatively, the grid-lines map onto both the distribution of points for the population as a whole, and the frequency-dependent evolution marked out for individual pulsars. We see that for $p \gtrsim 0.3$ the frequency evolution is predominantly horizontal, with $\theta \lesssim 20\degree$, whereas for $p \lesssim 0.3$ the frequency evolution is either diagonal or vertical. This behaviour is most concisely explained as resulting from frequency evolution of the ratio of mode strengths $R$, with a small amount of coherent mode addition taking place. It would be insufficient to have purely incoherent mode addition, even with elliptical modes, as it is impossible for incoherent addition alone to result in the observed diagonal or vertical tracks on the $p$--$\theta$ diagram.

\section{Individual pulsars and the partial-coherence model}
\label{sec:indivpsrs}

The three parameters of the partial-coherence model can be combined in a large number of different ways, which has the potential to lead to considerable complexity in observations. However, it is instructive to consider what happens to observable polarization when two parameters are held constant and a third is varied smoothly with either pulse phase or frequency. This provides both a clear example of model behaviour and evidence in favour of the partial-coherence model successfully explaining the observed polarization complexity in a simple way. Motivated by the population-level polarization behaviour of Fig. \ref{fig:RetaCpthetaDATA}, we searched for ordered polarization tracks in $p$--$\theta$ space for individual pulsars. We present here some individual examples of pulsars for which just such a scenario can neatly explain the observed polarization behaviour.

\subsection{Inferring model parameters from observed Stokes parameters}

In order to investigate individual pulsars, we first need to consider precisely how the mathematics of the model can be related to real observations. The three parameters of the partial-coherence model may be used to generate normalized model-Stokes parameters with the following formulae, derived in Appendix \ref{app:maths}:
\begin{equation}
\begin{split}
    &I_{m} = \left((1-C)^{2} + C^{2}\right)\left(1 + R\right) \\
    &Q_{m} = \left((1-C)^{2} + C^{2}\right)\left(1 - R\right) \\
    &U_{m} = 2\sqrt{R}C^{2}\cos{\eta} \\
    &V_{m} = 2\sqrt{R}C^{2}\sin{\eta} \\
\end{split}
\label{eq:Stokessummary}
\end{equation}

These parameters are normalized and derived with respect to a particular set of axes. They are therefore not directly comparable to the observational Stokes parameters $I_{d}, Q_{d}, U_{d}$ and $V_{d}$, where the subscripts $d$ and $m$ refer to data and model respectively. However, if we know the geometry of a pulsar, then the RVM defines the phase-dependent basis underlying the geometrical part of the PA. The model Stokes parameters then describe deviations away from this basis that result from the coherent combination of the modes, explaining the simultaneous presence of deviations from the RVM in the PA profile and the presence of circular polarization. 

If we do not know the pulsar geometry, it is still possible to make use of the model without reference to the parameter basis for a given pulsar. We can consider only the absolute fractions of linear and circular polarization, and neglect polarization angle and circular polarization handedness. Doing so results in two key limitations on the inferred model parameters. We can infer the ratio of mode strengths, $R$, but not which mode is stronger, so $0\leq R\leq 1$ always. For example, if one mode is twice the size of the other, but we have no information about which is which, by default the maths will infer $R = 0.5$ rather than $R = 2$. We can also infer the size of the phase offset, but not the basis from which that is measured, so $0\degree\leq\eta\leq90\degree$ always. For example, in terms of the relative magnitudes of linear and circular polarization, a phase offset $\eta = 120\degree$ is equivalent to a phase offset $\eta = 180\degree - 120\degree = 60\degree$, and by default the maths will infer $\eta = 60\degree$. Care must therefore be taken to recognise these limitations and, if possible, account for them with additional information.

This basis-free inference of partial-coherence model parameters is done as follows. Unlike the model Stokes parameters, fractional linear polarization and fractional absolute circular polarization are directly comparable for the data and the model: 
\begin{equation}
    \begin{split}
        l &= L_{d}/I_{d} = L_{m}/I_{m} = \sqrt{Q_{m}^{2} + U_{m}^{2}}/I_{m} \\
        \lvert v\rvert &= \lvert V_{d}\rvert/I_{d} = \lvert V_{m}\rvert/I_{m}, \\
    \end{split}
\label{eq:l_and_v}
\end{equation}
since these do not require the angle of linear polarization or the absolute mode strength. We have already described mapping the model parameters $R$, $\eta$ and $C$ onto these polarization fractions in Section \ref{sec:map}. Now we are seeking to move in the opposite direction and infer the partial-coherence model parameters from our polarization observations.

We can convert observed Stokes parameters ($I_{d}, Q_{d}, U_{d}$, $V_{d}$) to model parameters ($R$, $\eta$, $C$) by combining and rearranging equation blocks \ref{eq:Stokessummary} and \ref{eq:l_and_v}. In this set-up, we have two input data parameters ($l$ and $\lvert v\rvert$) and three output parameters ($R$, $\eta$ and $C$). This means that it is only possible to invert the model if we make an assumption about the value of one of the model parameters. Here, as an example, we assume we know the value of $\eta$ so that we can infer $R$ and $C$. We present the equations here and describe their derivation in detail in Appendices \ref{app:appendixgamma} and \ref{app:appendixC}:

\begin{equation}
  R=\begin{cases}
    \frac{1-l}{1+l}, & \text{if $\eta = 90\degree$},\\
    \frac{1-\sqrt{l^{2} - \left( \frac{\lvert v\rvert}{\tan\eta}\right)^{2}}}{1+\sqrt{l^{2} - \left( \frac{\lvert v\rvert}{\tan\eta}\right)^{2}}}, & \text{otherwise},
  \end{cases}
  \label{eq:Rsummary}
\end{equation}

\begin{equation}
  C=\begin{cases}
    \frac{1}{2}, & \text{if $2\lvert v\rvert = \lvert \alpha\rvert$},\\
    \frac{\lvert v\rvert - \sqrt{\lvert v\rvert(\lvert \alpha\rvert-\lvert v\rvert)}}{2\lvert v\rvert - \lvert \alpha\rvert}, & \text{otherwise},
  \end{cases}
  \label{eq:Csummary}
\end{equation}
where $\alpha~=~\sqrt{(1-l^{2})\sin^{2}{\eta} + \lvert v\rvert^{2}\cos^{2}{\eta}}$.

For a given pair of $l$ and $\lvert v\rvert$ we can infer the upper and lower bounds on the model parameters in the following way. Without information about the mode basis provided by the pulsar geometry, the maximum possible value of $\eta$ will always be $\eta_{\rm{max}}=90\degree$, as this means that all coherent addition will generate circular polarization only. The minimum possible value inferred value of $\eta$ is given by $\eta_{\rm{min}}=\theta=\arctan{(\lvert v\rvert/l)}$. A smaller value of $\eta$ would not be able to generate the amount of circular polarization measured. Note that this assumes that $0\degree\leq\eta\leq 90\degree$: the symmetry of $\eta$ about $90\degree$ means that the inferred $\eta_{\rm{min}}$ could equivalently result from $\eta_{\rm{true}} = 180\degree - \eta_{\rm{min}}$. With this in mind, for a given $\eta_{\rm{max}}$ and $\eta_{\rm{min}}$ we can then infer $R(\eta_{\rm{max}})$, $R(\eta_{\rm{min}})$, $C(\eta_{\rm{max}})$ and $C(\eta_{\rm{min}})$. This is the technique we employ in the following subsections to investigate the polarization behaviour of PSRs~J0820$-$1350, J1157$-$6224 and J0134$-$2937. For this work we use observations made with Murriyang, the Parkes radio telescope, as described in \cite{Oswald2023}.

\subsection{PSR~J0820$-$1350: frequency-dependence of mode phase offset}

\begin{figure*}
    \centering
    \includegraphics[width=\textwidth]{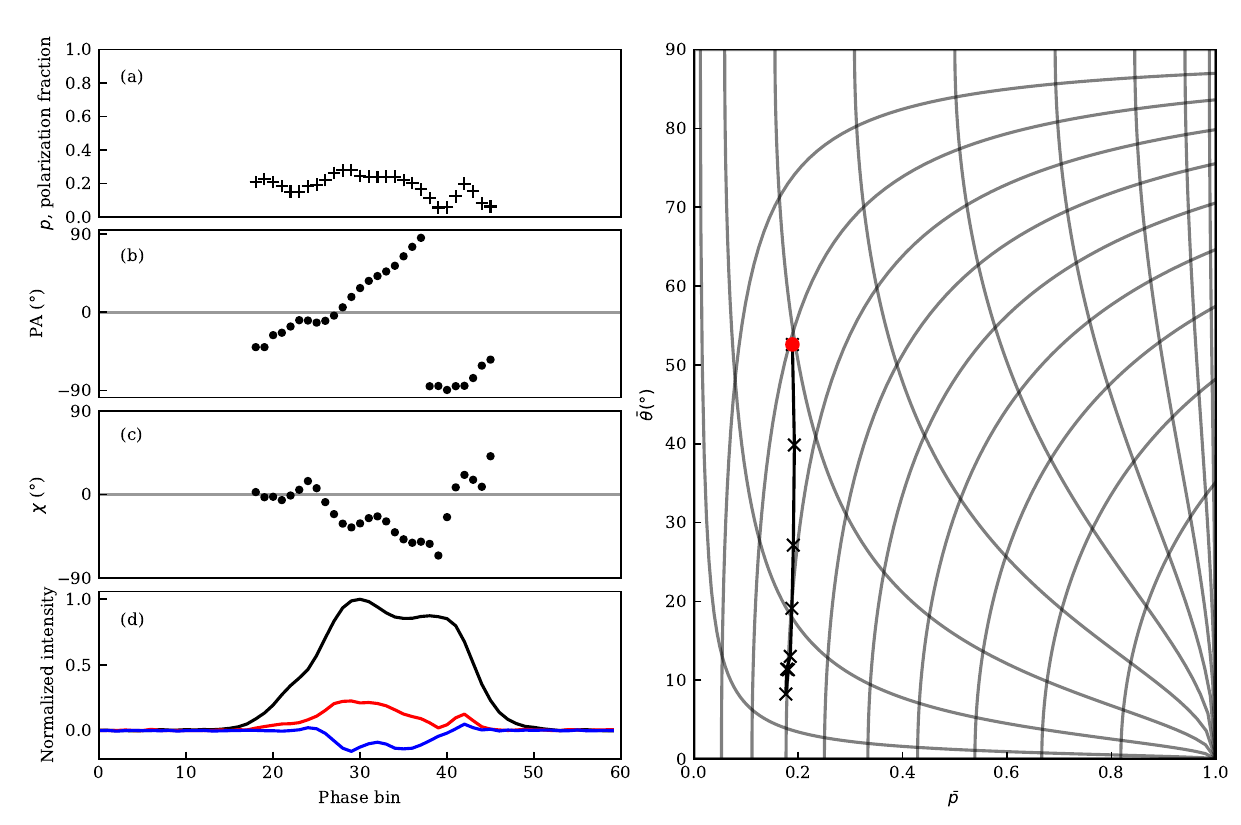}
    \caption{Frequency-dependent polarization of PSR~J0820$-$1350. Left: pulse profile properties at 1400~MHz: (a) polarization as a fraction of total intensity, $p$, (b) position angle, (c) ellipticity, (d) total intensity (black), linear polarization (red) and circular polarization (blue) all in normalized units. The variables in subplots (a), (b) and (c) are shown only where $p < 1$ and $p > 3\sigma_{I}$, where $\sigma_{I}$ is the standard deviation of the off-pulse region of the total intensity pulse profile. Right: $\bar{p}$--$\bar{\theta}$ diagram showing how the phase-averaged parameters vary with frequency between 809~MHz and 3782~MHz, with gridlines of varying $R$ and $C$ over-plotted. Points from successive frequencies are linked by lines and the value for the lowest frequency is marked with a red circle.}
    \label{fig:J0820}
\end{figure*}

Building on the context of Fig. \ref{fig:RetaCpthetaDATA}, we begin by investigating the frequency-dependent polarization behaviour of an individual pulsar. In Fig. \ref{fig:J0820} we show the pulse profile of PSR~J0820$-$1350 at 1400~MHz (left) and how the phase-averaged polarization values $\bar{p}$ and $\bar{\theta}$ vary with observing frequency (right). Note that, unlike the definitions given in equations \ref{eq:p_intro} and \ref{eq:theta_intro} we are considering phase-averaged parameters here: $\bar{p}=\Sigma(\sqrt{Q^{2} + U^{2} + V^{2}}^{BC})/\Sigma I$ and $\bar{\theta}=\arctan{(\Sigma(\lvert V\rvert)^{BC}/\Sigma(\sqrt{Q^{2} + U^{2}})^{BC})}$ where $\Sigma$ indicates a sum across pulse phase and the superscript $BC$ indicates that the polarization variables have been bias-corrected according to the prescriptions of \cite{Everett2002}. We see that the frequency-dependent average polarization forms a vertical track on the $\bar{p}$-$\bar{\theta}$ diagram: as we move from low to high frequency circular polarization is being converted to linear polarization with total polarization fraction staying constant. From Fig. \ref{fig:RetaCptheta} (top-middle and top-right subplots) we see that such a vertical track can be conceptualized as a smooth variation of mode phase offset $\eta$ with $R$ and $C$ held constant. It should be noted that since we are using phase-averaged inputs, there is likely to be a loss of information associated with the phase-averaging when we translate to the model parameters: we label them $R'$, $\eta'$ and $C'$ to emphasise this. We investigate how well such a picture describes the observed polarization.

We model the average polarization of PSR~J0820$-$1350 as follows. First, since we are assuming that all frequency-dependent polarization changes result only from changes of $\eta$, this means the phase-averaged polarization fraction $\bar{p}$ is constant across frequency. We set it equal to the mean of the polarization fraction measurements: $\mu$($p$) = 0.185. These measurements have a standard deviation of $\sigma$($p$) = 0.006, so this is a reasonable assumption in this case.

Next, we wish to identify an observing frequency at which $\bar{\theta} = 90\degree$: this would constrain $\eta' = 90\degree$, allowing us to infer the model parameters $R'$ and $C'$. We could then fix $R'$ and $C'$ and hence infer $\eta'$ at every other frequency. However, it can be seen from Fig. \ref{fig:J0820} that none of our measurements of $\bar{\theta}$ lie close to $90\degree$. We therefore look to the Poincar\'e sphere for additional information to help constrain $\eta'$.

\begin{figure}
    \centering
    \includegraphics[width=\columnwidth]{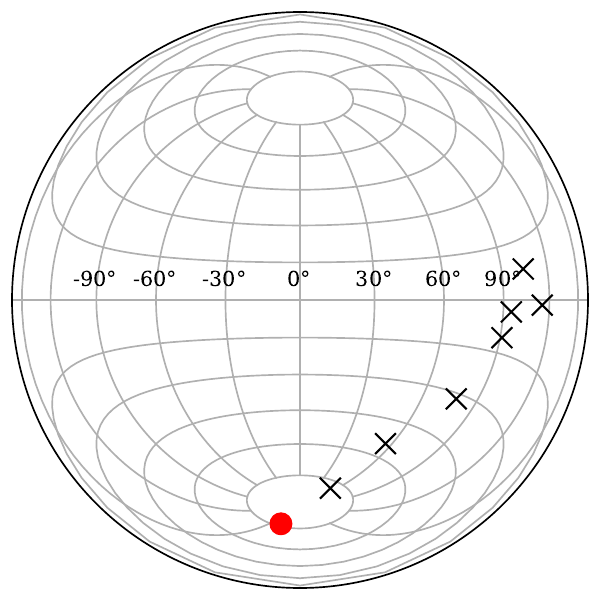}
    \caption{Lambert projection of the Poincar\'e sphere, showing the phase-averaged Stokes parameters of PSR~J0820$-$1350 at eight successive frequencies. As in Fig. \ref{fig:J0820}, the lowest frequency is marked with a red circle.}
    \label{fig:J0820_poincaresphere}
\end{figure}

In Fig. \ref{fig:J0820_poincaresphere}, we show the Stokes parameters $Q$, $U$ and $V$ on the Poincar\'e sphere, averaged across pulse phase and normalized by averaged total intensity. It is important to note that, because we are phase-averaging, the values on the Poincar\'e sphere in Fig. \ref{fig:J0820_poincaresphere} are not the same as the values in the $\bar{p}$--$\bar{\theta}$ plot in Fig. \ref{fig:J0820}. For the former, we are summing Stokes parameters $Q$, $U$ and $V$ across pulse phase, whereas to generate the parameters $\bar{p}$ and $\bar{\theta}$ we use the phase-summed total polarization, linear polarization and absolute circular polarization.

Nevertheless, we can use the Poincar\'e sphere to give a general indication of the angular dependence of the polarization state with frequency. We can see that in Fig. \ref{fig:J0820_poincaresphere} the lowest frequency polarization state (red circle) lies on the other side of the negative Stokes $V$ pole from the rest of the polarization states at higher frequencies (black crosses). From $\bar{\theta}$ alone, we know only that $\bar{\theta}\leq\eta'\leq 180\degree-\bar{\theta}$ at a given frequency. From the Poincar\'e sphere, we can constrain that if $\eta' > 90\degree$ at the lowest frequency, then $\eta' < 90\degree$ at the higher frequencies. Alternatively, if $\eta' < 90\degree$ at the lowest frequency, then $\eta' > 90\degree$ at the higher frequencies.

The frequency, $\nu$, at which $\eta'$ is most tightly constrained is the lowest frequency, because at this frequency the pulsar has its largest value of $\bar{\theta}$. Calling that frequency $\nu_{0}$, we know that $\bar{\theta}(\nu_{0})\leq\eta'(\nu_{0})\leq 180\degree-\bar{\theta}(\nu_{0})$. We have also inferred that $\eta' = 90\degree$ lies between $\nu_{0}$ and the next-lowest frequency: this means either that $\eta'(\nu_{0}) \geq 90\degree$ and $\eta'(\nu\neq\nu_{0}) < 90\degree$, or vice versa. We choose the former with no loss of generality.

We therefore have a defined range of $\eta'$ at the lowest frequency: $90\degree \leq \eta'(\nu_{0}) \leq 180\degree - \bar{\theta}(\nu_{0})$. This gives us limits on $R'$ and $C'$ at this frequency as follows:
\begin{align}
    \eta'_{\rm{min}}(\nu_{0}) &= 90\degree, \\
    R'(\eta'_{\rm{min}}(\nu_{0}), \nu_{0}) &= 0.80(\pm0.01), \\
    C'(\eta'_{\rm{min}}(\nu_{0}), \nu_{0}) &= 0.294(\pm0.004)
\end{align}
and 
\begin{align}
    \eta'_{\rm{max}}(\nu_{0}) &= 180\degree - \bar{\theta}(\nu_{0}) = 127\degree\pm3\degree, \\
    R'(\eta'_{\rm{max}}(\nu_{0}), \nu_{0}) &= 1.0(+0.0-0.1), \\
    C'(\eta'_{\rm{max}}(\nu_{0}), \nu_{0}) &= 0.32(+0.00-0.01), 
\end{align}
respectively. Fixing $R'$ and $C'$ as constants, we can then calculate $\eta'(\nu)$ as a function of frequency, being careful to choose $\eta'(\nu) < 90\degree$ at all other frequencies. 

\begin{figure*}
    \centering
    \includegraphics[width=\textwidth]{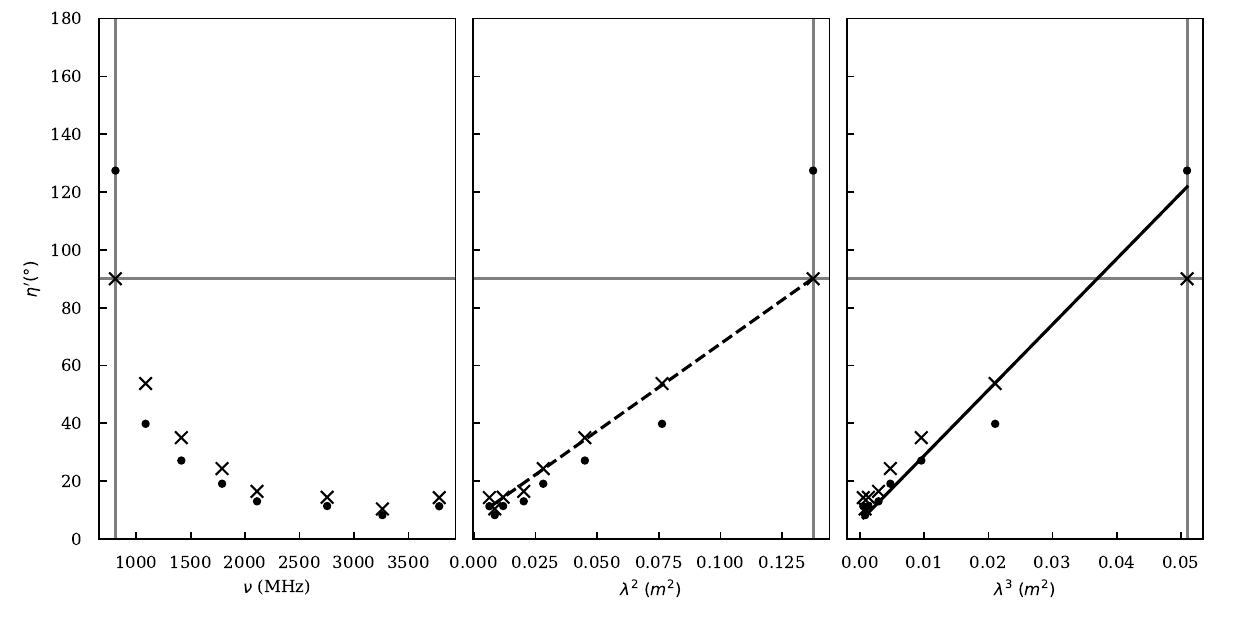}
    \caption{Modelled values for frequency-dependent $\eta'$ for PSR~J0820$-$1350, shown plotted against frequency $\nu$ (left), wavelength-squared $\lambda^{2}$ (middle) and wavelength-cubed $\lambda^{3}$ (right). The vertical line marks $\nu_{0}$ (or equivalently $\lambda^{2}_{0}$ or $\lambda^{3}_{0}$), the frequency at which $\eta'(\nu_{0})$ is fixed at either $\eta'_{\rm{min}}(\nu_{0}) = 90\degree$ (cross) or $\eta'_{\rm{max}}(\nu_{0}) = 180\degree - \theta(\nu_{0}) = 127\degree$ (point). The values of $\eta'$ at every other phase are then inferred for constant $R'$ and $C'$, as described in the text, and plotted with crosses and points. Two best fit straight lines are overplotted: the dashed line in the middle subplot is the best fit to the crosses of a linear relationship with $\lambda^{2}$ and the solid line in the right subplot is the best fit to the points of a linear relationship with $\lambda^{3}$. }
    \label{fig:J0820_eta_linear}
\end{figure*}

We find that $\eta'$ changes quickly at low frequency and more slowly at higher frequency, as shown in the first subplot of Fig. \ref{fig:J0820_eta_linear}. If we set $\eta'(\nu_{0}) = 90\degree$, the trend can be described as varying linearly with wavelength-squared ($\lambda^{2}$) and if we set $\eta'(\nu_{0}) = 180\degree - \bar{\theta}(\nu_{0}) = 127\degree$, the trend is better described as varying linearly with $\lambda^{3}$. We show these two trends in the second and third subplots of Fig. \ref{fig:J0820_eta_linear}.

In summary then, switching to wavelength-dependence rather than frequency-dependence, for $\eta'_{\rm{min}}(\lambda_{0})  = 90\degree$ we have
\begin{align}
    R' &= 0.80(\pm0.01), \\
    C' &= 0.294(\pm0.004), \\
    \eta'(\lambda) &= (604\degree/m^{2}\pm17\degree/m^{2})\lambda^{2} + (7\degree\pm1\degree).
\label{eq:linearEtaEq1}
\end{align}
where $\eta'$ is measured in degrees and $\lambda$ in metres. Similarly, for $\eta'_{\rm{max}}(\lambda_{0}) = 180\degree - \bar{\theta}(\lambda_{0}) = 127\degree$ we have
\begin{align}
    R' &= 1.0(+0.0-0.1), \\
    C' &= 0.32(+0.00-0.01), \\
    \eta'(\lambda) &= (2268\degree/m^{3}\pm142\degree/m^{3})\lambda^{3} + (6\degree\pm3\degree).
\label{eq:linearEtaEq2}
\end{align}

\subsection{PSR~J1157$-$6224: pulse-phase-dependence of mode phase offset}

\begin{figure*}
    \centering
    \includegraphics[width=\textwidth]{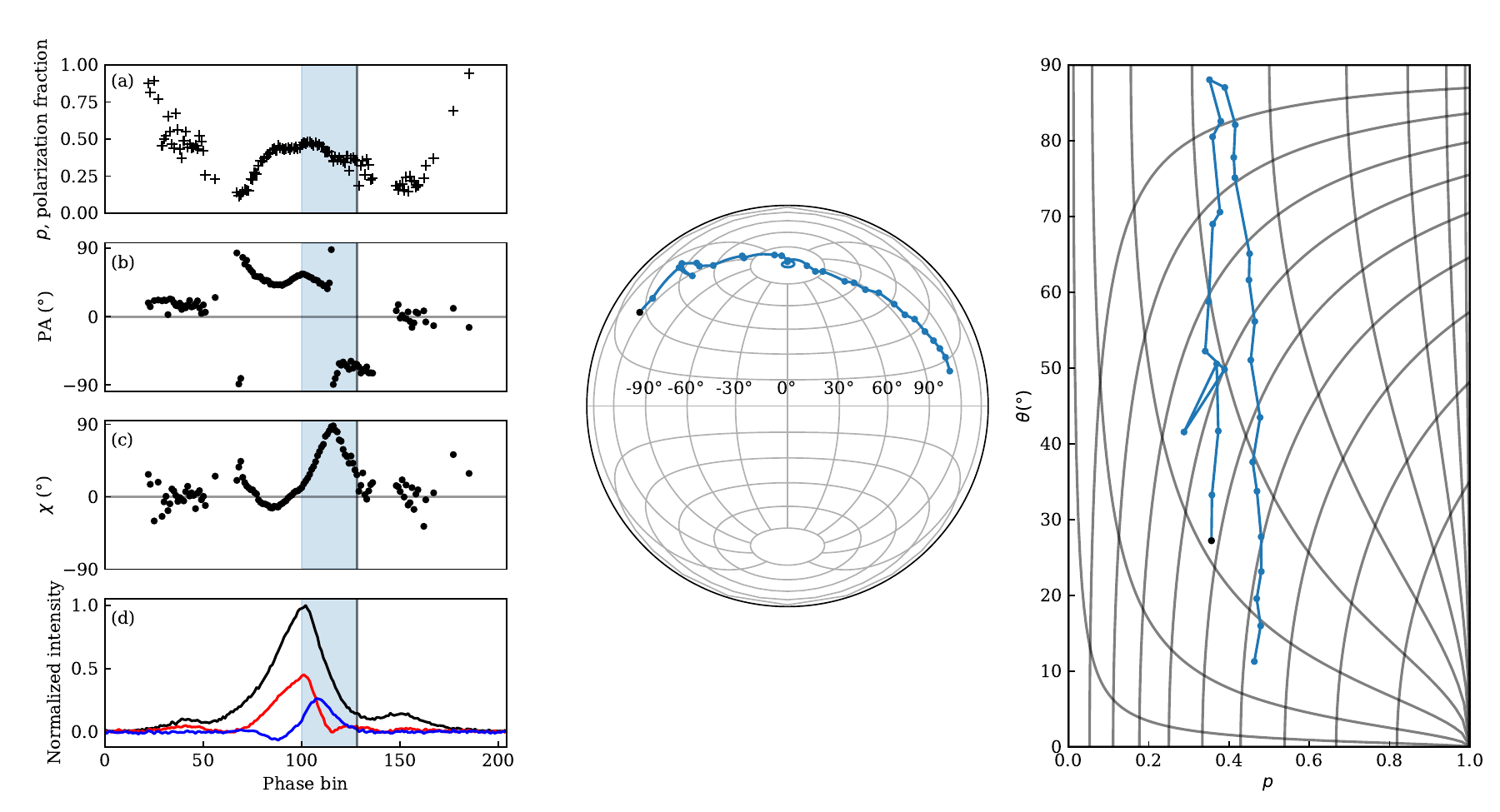}
    \caption{Phase-dependent polarization of PSR~J1157$-$6224. Left: pulse profile properties at 1400~MHz as described in Fig. \ref{fig:J0820}. Shaded blue phase region identifies the phase bins plotted on the middle and left subplots. Middle: Lambert projection of the Poincar\'e sphere, showing Stokes parameters for the blue-shaded phase region of the pulse profile. Points from successive phase bins are linked by lines and the value for the highest phase bin is marked in black: this corresponds to the phase bin marked by a vertical black line in plots (a) to (d). Right: $p$--$\theta$ diagram of the blue-shaded phase region, with gridlines of varying $R$ and $C$ over-plotted. As for the Poincar\'e sphere, points from successive phase bins are linked by lines and the value for the highest phase bin is marked in black.}
    \label{fig:J1157}
\end{figure*}

In addition to the frequency-dependent polarization properties described in \cite{Oswald2023} and shown in Fig. \ref{fig:RetaCpthetaDATA}, we witnessed many individual cases in our pulsar sample where polarization evolves with pulse phase, such that total polarization fraction stays near-constant, but the linear and circular fractions vary. A clear example of this is the phase-dependent linear and circular polarization of PSR~J1157$-$6224. As shown on the left of Fig. \ref{fig:J1157}, in the right half of the central component (shaded blue region) the linear polarization decreases with increasing phase as the circular polarization increases, and then vice versa, such that the total polarization fraction remains near-constant. 
The phase-dependent polarization track is indicated on the Poincar\'e sphere in the middle of Fig. \ref{fig:J1157} and shows evidence of a transition of polarization state between linear and circular. Since in this case we are studying phase-resolved polarization, this means that, unlike the case of J0820$-$1350, there is a direct relationship between the values on the Poincar\'e sphere and those on the $p$--$\theta$ diagram. 
The polarization behaviour of this region of the pulse profile of PSR~J1157$-$6224 can be visualized as a vertical track on the $p$-$\theta$ diagram that rises and falls, as shown on the right of Fig. \ref{fig:J1157}. A vertical track once more suggests that $\eta$ is changing whilst $R$ and $C$ are held constant.

We model this phase-region of PSR~J1157$-$6224 as follows. First, just as for J0820$-$1350, we require the polarization fraction $p$ to be constant across pulse phase. We use $\mu(p) = 0.41 \pm 0.05$. Next, we calculate the model parameters $R$, $C$ and $\eta$ for the pulse phase with the highest value of $\theta$, as it is at this phase bin that the parameters will be most constrained. Calling this phase bin $\phi_{0}$, we infer the following parameters: 
\begin{align}
    \eta_{\rm{min}}(\phi_{0}) &= \theta(\phi_{0}) = 88\degree\pm3\degree \\
    & (\equiv 180\degree - \theta(\phi_{0}) = 92\degree\pm 3\degree), \\
    R(\eta_{\rm{min}}(\phi_{0}), \phi_{0}) &= 1.00(+0.00-0.06), \\
    C(\eta_{\rm{min}}(\phi_{0}), \phi_{0}) &= 0.4543(+0.0000-0.0005), 
\end{align}
as one set of constraints. Similarly, we calculate the model parameters for $\eta_{\rm{max}} = 90\degree$, giving 
\begin{align}
    \eta_{\rm{max}}(\phi_{0}) &= 90\degree, \\
    R(\eta_{\rm{max}}(\phi_{0}), \phi_{0}) &= 0.97(\pm0.03), \\
    C(\eta_{\rm{max}}(\phi_{0}), \phi_{0}) &= 0.4542(+0.0001-0.0005)
\end{align}
as the other set. The error on $\eta_{\rm{min}}(\phi_{0})$ is given as the error on $\theta(\phi_{0})$, which is inferred from the standard deviation of the off-pulse region of the Stokes parameters. Varying the input parameters $\theta$ and $\phi$ by this amount changes $R$ and $C$ by a small amount, which gives the precisions to which we have quoted the values above.

Next, we make the assumption that $R$ and $C$ are constants and do not vary with pulse phase. This means we now have enough information to calculate $\eta(\phi)$ at every other pulse phase. We note that, when considering only the total linear and absolute circular polarization and not making use of the phase information, we are only able to calculate a value of $\eta$ between $0\degree$ and $90 \degree$, since every value of $\eta > 90\degree$ would lead to an equivalent ratio of linear and circular polarization as a value of $\eta$ within this range. However, from the Poincar\'e sphere in Fig. \ref{fig:J1157} we see that the circular polarization increases to a maximum and then decreases again, such that the linear polarization continues to rotate around the sphere. It is therefore logical to infer from this that the value of $\eta$ actually increases beyond $90\degree$ at the higher phases (or indeed the mirror image - it could start out greater than $90\degree$ and decrease with phase.).

\begin{figure}
    \centering
    \includegraphics[width=\columnwidth]{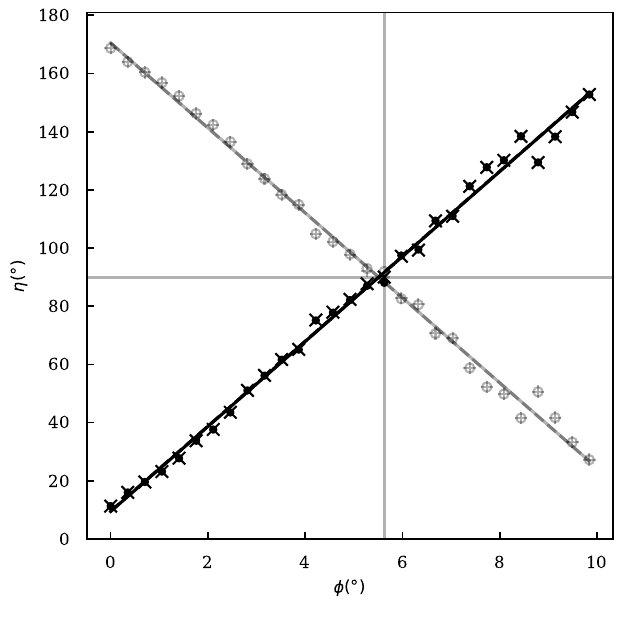}
    \caption{Modelled values for phase-dependent $\eta$ for PSR~J1157$-$6224. The vertical line marks $\phi_{0}$, where $\eta(\phi_{0})$ is fixed at either $\eta_{\rm{max}}(\phi_{0}) = 90\degree$ (cross) or $\eta_{\rm{min}}(\phi_{0}) = \theta(\phi_{0}) = 88\degree$ (point). The values of $\eta$ at every other phase are then inferred for constant $R$ and $C$ as described in the text and plotted with crosses and points. The best fit straight lines are overplotted: the dashed line is the best fit to the crosses and the solid line is the best fit to the points. Since $\eta$ is symmetrical about $90\degree$, it could be decreasing with phase rather than increasing: we plot the mirror image values of $\eta$ as plusses and open circles.}
    \label{fig:J1157_eta_linear}
\end{figure}

Taking this into account, we find that $\eta$ varies linearly with pulse phase. This is true for both limiting parameter sets ($\eta_{\rm{min}}$ and $\eta_{\rm{max}}$), as is shown in Fig. \ref{fig:J1157_eta_linear}. We therefore perform a least-squares fit for linear dependencies of $\eta$ on $\phi$ in both cases: the best fits are plotted in Fig. \ref{fig:J1157_eta_linear} and are very similar to each other, so this behaviour is tightly constrained.

In summary, the phase-dependent polarization behaviour for this region of the profile of PSR~J1157$-$6224 can be described by constant $R$, constant $C$ and phase-varying $\eta(\phi)$. By calculating the constraints on the initial choice of $\eta(\phi_{0})$, we infer the parameter ranges that describe the whole region. 

For $\eta_{\rm{min}}(\phi_{0}) = \theta(\phi_{0}) = 88\degree$ we have
\begin{align}
    R &= 1.00(+0.00-0.06), \\
    C &= 0.4543(+0.0000-0.0005), \\
    \eta(\phi) &= (14.6\pm0.2)\phi + (9\pm1)
\label{eq:linearEtaEq3}
\end{align}
where $\eta$ and $\phi$ are both measured in degrees. Note that the zero-point for rotational phase $\phi$ can be defined as any point in the pulsar's rotation and shifting the zero-point will alter the intercept but not the gradient of the fit to $\eta$. For simplicity we have defined $\phi = 0\degree$ to be the first phase bin of the phase region being investigated. Equivalently, $\eta$ could have a negative gradient, at which point the intercept $i$ would be given by $180\degree - i$ instead.

Similarly, for $\eta_{\rm{max}}(\phi_{0}) = 90\degree$ we have
\begin{align}
    R &= 0.97(\pm0.03), \\
    C &= 0.4542(+0.0001-0.0005), \\
    \eta(\phi) &= (14.6\pm0.2)\phi + (10\pm1).
\label{eq:linearEtaEq4}
\end{align}

\subsection{PSR~J0134$-$2937: pulse-phase-dependence of mode strength ratio}

\begin{figure*}
    \centering
    \includegraphics[width=\textwidth]{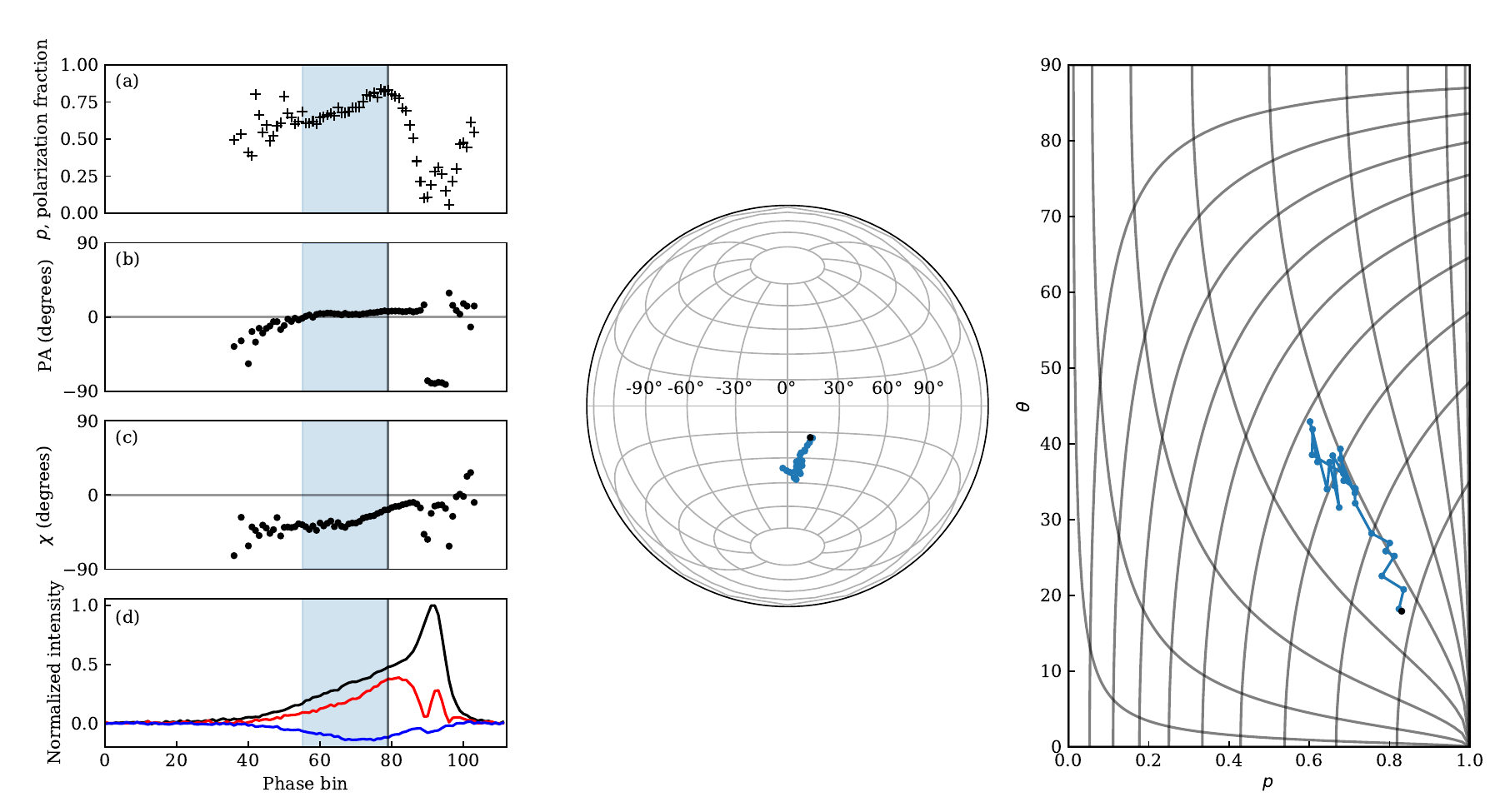}
    \caption{Figure set up as for Fig. \ref{fig:J1157} but now showing PSR~J0134$-$2937.
    }
    \label{fig:J0134_profileetc}
\end{figure*}

Finally, we consider the case of PSR~J0134$-$2937. As can be seen from Fig. \ref{fig:J0134_profileetc}, the selected phase region of the pulsar follows a diagonal path on the $p$-$\theta$ diagram with a negative slope, unlike the behaviour of PSR~J1157$-$6224. We test whether we can model this behaviour as resulting from a smooth variation of mode strength ratio $R$, with $C$ and $\eta$ held constant.

Following a similar method as for the previous case, we identify that $\eta$ (now a constant) must be at least as large as the largest value of $\theta$ for this phase region, or it will not be possible to use the same value of $\eta$ to describe the whole phase region. We obtain 
\begin{align}
    \eta_{\rm{min}} &= \theta(\phi_{0}) = 43\pm3\degree (\equiv 180\degree - \theta(\phi_{0}) = 137\degree\pm 3\degree), \\
    \eta_{\rm{max}} &= 90\degree.
\end{align}

\begin{figure}
    \centering
    \includegraphics[width=\columnwidth]{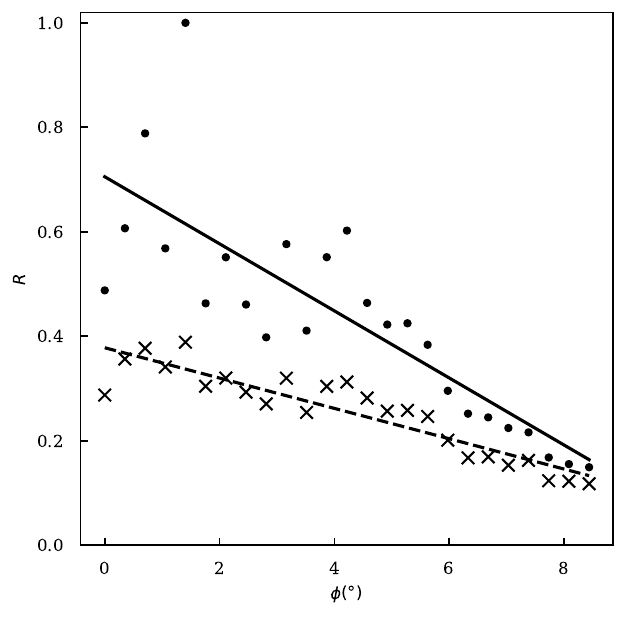}
    \caption{Modelled values for phase-dependent $R$ for PSR~J0134$-$2937. The values are shown for both $\eta_{\rm{max}} = 90\degree$ (crosses) and $\eta_{\rm{min}} = \theta(\phi_{0}) = 43\degree$ (points). The best fit straight lines are overplotted: the dashed line is the best fit to the crosses and the solid line is the best fit to the points.}
    \label{fig:J0134_R_linear}
\end{figure}

Now we hold $\eta$ constant across pulse phase and calculate both $R(\phi)$ and $C(\phi)$ as functions of phase. Doing so, we find that for $\eta_{\rm{max}} = 90\degree$, the polarization is well modelled with constant $C$ and linearly varying $R(\phi)$, as shown in Fig. \ref{fig:J0134_R_linear}. When $\eta_{\rm{min}} = \theta(\phi_{0}) = 43\degree$, $C$ is still well modelled as a constant but $R$ is somewhat less well constrained as varying linearly with $\phi$ (see Fig. \ref{fig:J0134_R_linear}). This may be evidence towards the true value of $\eta$ being at or close to $90\degree$. 

In summary, we find that for $\eta_{\rm{min}}~=~\theta(\phi_{0})~=~43\degree(\equiv 137\degree)$ we have
\begin{align}
    R(\phi) &= (-0.064\pm0.009)\phi + (0.71\pm0.05), \\
    C &= 0.56\pm0.02, 
\label{eq:linearEtaEq5}
\end{align}
For $\eta_{\rm{max}} = 90\degree$ we obtain
\begin{align}
    R(\phi) &= (-0.029\pm0.003)\phi + (0.38\pm0.01), \\
    C &= 0.48\pm0.02.
\label{eq:linearEtaEq6}
\end{align}

\section{Discussion}
\label{sec:discussion}

\subsection{Applicability of model to observations}

The partial-coherence model can explain many of the key observable features of pulsar polarization that relate to circular polarization. Whereas the concept of varying mode ratio has been discussed many times before in the context of orthogonal jumps, the inclusion of coherent addition into the model via parameters $C$ and $\eta$ enables us to explain more complex polarization phenomena under the same picture. It successfully explains the co-occurrence of deviations from the RVM in the PA with features in circular polarization as a rotation from Stokes $U$ into Stokes $V$; it can explain the frequency- and phase-dependence of polarization features in individual pulsars as resulting from the evolution of the model parameters in predictable ways; and it can be used to explain the trends in relation to polarization fraction and circular contribution, both across the population and across frequency for individual pulsars. 

The three pulsars we investigate in this paper show organised polarization behaviour for which our model reduces the number of degrees of freedom required to describe the polarization, in comparison to using Stokes parameters. Each pulsar exhibits polarization evolution behaviour that can be fully described by holding two of the partial-coherence model parameters constant and allowing the third to vary in a straightforward way.  These individual cases show variation of $R$ and $\eta$ with phase and $\eta$ with frequency, and it seems likely that other pulsars may exhibit other parameter variations.

We constrained the parameters based on upper and lower bounds on $\eta$. It is interesting to note that in the cases of PSRs~J0134$-$2937 and J0820$-$1350 the best fit solution was that for which $C$ was smaller (J1157$-$6224 is tightly constrained and the difference between its upper and lower bounds on $C$ is negligible). This tendency towards smaller $C$ is perhaps reflected on the population scale: it can be seen in Fig. \ref{fig:RetaCpthetaDATA} that there exist no $p$--$\theta$ measurements in the top-right corner of the plot, which corresponds to no measurements at a large value of $C$. We further note that the polarization measurements we use in this paper are all taken from the incoherent sum of single pulses to produce an integrated pulse profile, and that the frequency-dependent polarization measurements shown in Fig. \ref{fig:RetaCpthetaDATA} are also phase-averaged as well. This incoherent summing may result in a reduction of the magnitude of the measured coherence fraction $C$ compared to that which might perhaps be measured from individual (bright) single pulses.

A strength of the partial-coherence model is that it presupposes nothing about pulsar magnetosphere properties in order to model polarization and make predictions. Indeed, the mathematics of the model are simply a reparametrization of the Stokes parameters into the three-parameter collection $R$, $\eta$ and $C$, giving us a way to reconceptualize the relationship between linear and circular polarization and the pulse profile features we observe. From this perspective, the exact physical process by which Stokes $U$ is converted to Stokes $V$ is unconstrained, and the focus is more on understanding the frequency-dependent behaviour of relative mode intensities, and the ways in which the PA is rotated away from the underlying geometrical profile shape. As demonstrated with the examples in Section \ref{sec:indivpsrs}, the partial-coherence model can provide a simplistic description of what appear to be quite complex polarization phenomena when viewed purely in terms of Stokes parameters.

This paper has focused on polarization fractions $l$ and $\lvert v\rvert$, rather than Stokes parameters $Q$, $U$ and $V$, which depend on the choice of basis. However, the partial-coherence model can also be successfully applied to the full basis-dependent picture of pulsar polarization, where we include the information about the angle of linear polarization and sign of circular polarization. Future work will involve simultaneously fitting the PA profile for both the RVM and for the deviations resulting from the partially-coherent mode addition. Accounting for the RVM-deviations, circular polarization and frequency evolution with the partial-coherence model would enable improved measurements of pulsar geometries, building our understanding of the spin-evolution of pulsars across their lifetimes.

\subsection{Physical motivation for the partial-coherence model}
\label{sec:history}

Although this model successfully explains many observational polarization effects in the pulsar population, both collectively and for individual cases, there remain the questions of why the orthogonal modes for a given pulsar have a particular set of parameter values, and what the reasons are for those values evolving with phase, frequency and spin-down energy $\dot{E}$. We therefore briefly review the historical development of theories related to orthogonal mode interaction in the pulsar magnetosphere to provide some physical context for our model. Further information is summarized in the recent review of pulsar magnetospheres and their radiation by \cite{Philippov2022}.

The key observational factors that theory has sought to explain are the following: change in mode dominance with pulse phase; generation of circular polarization; frequency dependence of polarization behaviour; and pulsar behaviour, including polarization, varying with $\dot{E}$. The key theoretical concepts explored to explain these observations are generalized Faraday rotation, mode coupling, refraction and the effect of different heights of radio emission.

Most theoretical approaches to considering pulsar polarization consider the natural wave modes of the pulsar plasma to be linearly polarized, as that is more physically tractable \citep{Melrose1977}. We have further shown in this paper that allowing only incoherent combination of elliptically polarized orthogonal natural modes does not permit the frequency-evolution of circular contribution we observe. The circular polarization observed in pulse profiles must then be generated as a propagational effect. 

\cite{Kennett1998} described in detail how generalised Faraday rotation (GFR, also known as Faraday conversion) of pulsar radio emission could result from interaction with the medium through which the radio waves travel. This would result in the generation of circular polarization, with the effect being frequency-dependent. 
The case of PSR~J0820$-$1350 shows a rotation of polarization state from linear to circular in a way that can be modelled as depending on either wavelength-squared or wavelength-cubed. Wavelength-dependent conversion between linear and circular polarization could be explained by GFR, for which the exact wavelength relationship depends on the physical model invoked - for example \cite{Kennett1998} predict a $\lambda^{3}$ relationship, whereas the treatment of \cite{Lyutikov2022} predicts $\lambda^{\alpha}$ where $1\leq\alpha\leq 2$. Alternatively, \cite{Beniamini2022} showed that GFR effects could also originate through multipath propagation of radio waves through a magnetized scattering screen in the interstellar medium. Although this work focused on Fast Radio Bursts, they pointed out that such a mechanism would also frequency-dependent polarization features in pulsar observations. 

Circular polarization could also arise as a propagation effect involving mode coupling \citep[e.g.][]{Melrose1979}. \cite{Lyubarskii1998} and \cite{Petrova2001} developed the theoretical grounding for this picture, describing the effects of refraction, conversion between ordinary and extraordinary modes and mode coupling in the polarization limiting region as an origin for both circular polarization and non-orthogonal PA jumps.

Refraction in the magnetosphere would cause OPMs to travel along different paths, resulting in angular separation of the modes and affecting pulse widths, component separations and polarizations, including the observation of orthogonal jumps \citep{Barnard1986, Lyubarskii1998a, Petrova2000}. We note that angular separation of the modes, under the case of a changing magnetic geometry, could lead to the observed combination of non-orthogonal modes. Since our model provides only for orthogonal modes, this leads to an inherent limitation that we must make the assumption that this angular separation is small. \cite{VonHoensbroech1998} described the observational predictions of having two modes with different, frequency-dependent, refractive indices as being the following: 
\begin{enumerate}
    \item increasing frequency should be linked with decreasing linear and increasing circular polarization \citep[as also described by][]{Melrose1977}; 
    \item there should be a correlation between linear polarization and spin-down energy $\dot{E}$ at frequencies above a few GHz; 
    \item long period pulsars should have weaker linear polarization at high frequencies; 
    \item the difference between mode refractive indices should be smaller at higher frequencies, leading to depolarization. 
\end{enumerate}
All of these predictions have been verified in pulsar observations, for example \cite{Weltevrede2008} demonstrated the link between linear polarization and $\dot{E}$ at 1.5~GHz. In \cite{Oswald2023} we demonstrated that broad-band observations of a large sample of radio pulsars provide evidence for every one of these predictions. 

If the polarization behaviour we observed is strongly dependent on the propagation paths of the modes in the pulsar magnetosphere, as indicated by the theory described above, then the height of emission above the pulsar surface will be important in determining that polarization. \cite{Ruderman1975} described the theory of radius-to-frequency mapping to as a way of explaining the frequency-dependence of pulse widths: such a picture could also provide insight into the frequency-dependence of pulsar polarization. Another explanation is that presented by \cite{Karastergiou2005}, who proposed that frequency-dependent depolarization results from the orthogonal modes having different spectral indices. They added that this may also explain the differing polarization behaviour of young pulsars with high spin-down energy $\dot{E}$. The impact of differing emission heights may also affect how polarization is observed to vary with $\dot{E}$: \cite{Karastergiou2007} presented a model in which younger (high $\dot{E}$) and older (low $\dot{E}$) pulsars emit from different height ranges, leading to differing beam shapes and polarization behaviour.

Our model brings together the effects of mode propagation and interaction (affected by emission height, refraction and/or GFR) to result in observed partially-coherent polarization behaviour. Although the partial-coherence model makes no claims about the physical origins of polarization behaviour, the context of past research indicates that frequency-dependent behaviours could be linked to some combination of frequency-dependent mode strengths, refractive indices and emission heights and the $\dot{E}$-dependent behaviour could also be linked to differing emission height behaviour for low- and high-$\dot{E}$ pulsars. Disentangling the impacts of these different physical explanations may require clearer predictions about frequency-dependent behaviour and a closer look at the phase-resolved polarization of pulsar emission, and perhaps how it changes over time.

\section{Conclusions}
\label{sec:conc}

Motivated by the observations of broad-band radio pulsar polarization detailed in \cite{Oswald2023}, we develop the partial-coherence model to describe how the interaction of orthogonal modes of radio emission generates circular polarization in co-occurrence with other key polarization features: rotation of the PA away from the RVM, frequency evolution of polarization, pulse profile complexity and polarization fraction. The partial-coherence model is a simple mathematical description of linearly polarized orthogonal modes interacting to generate circular polarization and is fully described by three parameters: the mode strength ratio, mode phase offset and coherence fraction. Partial coherence is the clearest and easiest way to explain the diverse complexity of polarization behaviour seen in broadband radio pulsar profiles. Future work will extend the model description beyond the average-polarization-fraction set-up here to encompass the angle of linear polarization and sign of circular polarization, and use it to model deviations from the RVM in PA profiles to improve measurements of pulsar geometries.

\section*{Acknowledgements}
We would like to thank the reviewer for careful consideration of this manuscript and helpful advice. 
Murriyang, the Parkes radio telescope, is part of the Australia Telescope National Facility (\href{https://ror.org/05qajvd42}{https://ror.org/05qajvd42}) which is funded by the Australian Government for operation as a National Facility managed by CSIRO. 
We acknowledge the Wiradjuri people as the traditional owners of the Observatory site. 
LSO acknowledges the support of Magdalen College, Oxford.

\section*{Data Availability}
The data underlying this article will be shared on reasonable request to the corresponding author.

%%%%%%%%%%%%%%%%%%%%%%%%%%%%%%%%%%%%%%%%%%%%%%%%%%



\bibliographystyle{mnras}
\bibliography{PolarizationModel} % if your bibtex file is called example.bib

\begin{thebibliography}{}
\makeatletter
\relax
\def\mn@urlcharsother{\let\do\@makeother \do\$\do\&\do\#\do\^\do\_\do\%\do\~}
\def\mn@doi{\begingroup\mn@urlcharsother \@ifnextchar [ {\mn@doi@}
  {\mn@doi@[]}}
\def\mn@doi@[#1]#2{\def\@tempa{#1}\ifx\@tempa\@empty \href
  {http://dx.doi.org/#2} {doi:#2}\else \href {http://dx.doi.org/#2} {#1}\fi
  \endgroup}
\def\mn@eprint#1#2{\mn@eprint@#1:#2::\@nil}
\def\mn@eprint@arXiv#1{\href {http://arxiv.org/abs/#1} {{\tt arXiv:#1}}}
\def\mn@eprint@dblp#1{\href {http://dblp.uni-trier.de/rec/bibtex/#1.xml}
  {dblp:#1}}
\def\mn@eprint@#1:#2:#3:#4\@nil{\def\@tempa {#1}\def\@tempb {#2}\def\@tempc
  {#3}\ifx \@tempc \@empty \let \@tempc \@tempb \let \@tempb \@tempa \fi \ifx
  \@tempb \@empty \def\@tempb {arXiv}\fi \@ifundefined
  {mn@eprint@\@tempb}{\@tempb:\@tempc}{\expandafter \expandafter \csname
  mn@eprint@\@tempb\endcsname \expandafter{\@tempc}}}

\bibitem[\protect\citeauthoryear{Barnard \& Arons}{Barnard \&
  Arons}{1986}]{Barnard1986}
Barnard J.~J.,  Arons J.,  1986, \mn@doi [ApJ] {10.1086/163979}, 302, 138

\bibitem[\protect\citeauthoryear{Beniamini, Kumar  \& Narayan}{Beniamini
  et~al.}{2022}]{Beniamini2022}
Beniamini P.,  Kumar P.,   Narayan R.,  2022, \mn@doi [MNRAS]
  {10.1093/mnras/stab3730}, 510, 4654

\bibitem[\protect\citeauthoryear{Cordes, Rankin  \& Backer}{Cordes
  et~al.}{1978}]{CordesRankinBacker1978}
Cordes J.~M.,  Rankin J.,   Backer D.~C.,  1978, \mn@doi [ApJ]
  {10.1086/156328}, 223, 961

\bibitem[\protect\citeauthoryear{Desvignes et~al.,}{Desvignes
  et~al.}{2019}]{Desvignes2019b}
Desvignes G.,  et~al., 2019, \mn@doi [Science] {10.1126/science.aav7272}, 365,
  1013

\bibitem[\protect\citeauthoryear{Dyks}{Dyks}{2019}]{Dyks2019}
Dyks J.,  2019, \mn@doi [MNRAS] {10.1093/mnras/stz1690}, 488, 2018

\bibitem[\protect\citeauthoryear{Everett \& Weisberg}{Everett \&
  Weisberg}{2002}]{Everett2002}
Everett J.~E.,  Weisberg J.~M.,  2002, \mn@doi [ApJ] {10.1086/320652}, 553, 341

\bibitem[\protect\citeauthoryear{Gangadhara}{Gangadhara}{1997}]{Gangadhara1997}
Gangadhara R.~T.,  1997, A{\&}A, 327, 155

\bibitem[\protect\citeauthoryear{Johnston \& Kramer}{Johnston \&
  Kramer}{2019}]{Johnston2019a}
Johnston S.,  Kramer M.,  2019, \mn@doi [MNRAS] {10.1093/mnras/stz2865}, 490,
  4565

\bibitem[\protect\citeauthoryear{Johnston \& Weisberg}{Johnston \&
  Weisberg}{2006}]{Johnston2006}
Johnston S.,  Weisberg J.~M.,  2006, \mn@doi [MNRAS]
  {10.1111/j.1365-2966.2006.10263.x}, 368, 1856

\bibitem[\protect\citeauthoryear{Karastergiou \& Johnston}{Karastergiou \&
  Johnston}{2007}]{Karastergiou2007}
Karastergiou A.,  Johnston S.,  2007, \mn@doi [MNRAS]
  {10.1111/j.1365-2966.2007.12237.x}, 380, 1678

\bibitem[\protect\citeauthoryear{Karastergiou, Johnston  \&
  Manchester}{Karastergiou et~al.}{2005}]{Karastergiou2005}
Karastergiou A.,  Johnston S.,   Manchester R.~N.,  2005, \mn@doi [MNRAS]
  {10.1111/j.1365-2966.2005.08909.x}, 359, 481

\bibitem[\protect\citeauthoryear{Karastergiou, Roberts, Johnston, Lee,
  Weltevrede  \& Kramer}{Karastergiou et~al.}{2011}]{Karastergiou2011}
Karastergiou A.,  Roberts S.~J.,  Johnston S.,  Lee H.,  Weltevrede P.,
  Kramer M.,  2011, \mn@doi [MNRAS] {10.1111/j.1365-2966.2011.18697.x}, 415,
  251

\bibitem[\protect\citeauthoryear{Kennett \& Melrose}{Kennett \&
  Melrose}{1998}]{Kennett1998}
Kennett M.,  Melrose D.,  1998, \mn@doi [PASA] {10.1071/AS98211}, 15, 211

\bibitem[\protect\citeauthoryear{Lyubarskii \& Petrova}{Lyubarskii \&
  Petrova}{1998a}]{Lyubarskii1998}
Lyubarskii Y.~E.,  Petrova S.~A.,  1998a, \mn@doi [Astrophys. Space Sci.]
  {10.1023/A:1001872805645}, 262, 379

\bibitem[\protect\citeauthoryear{Lyubarskii \& Petrova}{Lyubarskii \&
  Petrova}{1998b}]{Lyubarskii1998a}
Lyubarskii Y.~E.,  Petrova S.~A.,  1998b, A{\&}A, 333, 181

\bibitem[\protect\citeauthoryear{Lyutikov}{Lyutikov}{2022}]{Lyutikov2022}
Lyutikov M.,  2022, \mn@doi [ApJ Lett.] {10.3847/2041-8213/ac786f}, 933, L6

\bibitem[\protect\citeauthoryear{Manchester, Taylor  \& Huguenin}{Manchester
  et~al.}{1975}]{Manchester1975}
Manchester R.~N.,  Taylor J.~H.,   Huguenin G.~R.,  1975, \mn@doi [ApJ]
  {10.1086/153395}, 196, 83

\bibitem[\protect\citeauthoryear{Melrose}{Melrose}{1979}]{Melrose1979}
Melrose D.~B.,  1979, \mn@doi [Aust. J. Phys.] {10.1071/PH790061}, 32, 61

\bibitem[\protect\citeauthoryear{Melrose \& Stoneham}{Melrose \&
  Stoneham}{1977}]{Melrose1977}
Melrose D.~B.,  Stoneham R.~J.,  1977, Proc. ASA, 3, 120

\bibitem[\protect\citeauthoryear{Michel}{Michel}{1987}]{Michel1987}
Michel F.~C.,  1987, \mn@doi [ApJ] {10.1086/165775}, 322, 822

\bibitem[\protect\citeauthoryear{Oswald et~al.,}{Oswald
  et~al.}{2023}]{Oswald2023}
Oswald L.~S.,  et~al., 2023, \mn@doi [MNRAS] {10.1093/mnras/stad070}, 520, 4961

\bibitem[\protect\citeauthoryear{Petrova}{Petrova}{2001}]{Petrova2001}
Petrova S.~A.,  2001, \mn@doi [A{\&}A] {10.1051/0004-6361:20011297}, 378, 883

\bibitem[\protect\citeauthoryear{Petrova \& Lyubarskii}{Petrova \&
  Lyubarskii}{2000}]{Petrova2000}
Petrova S.~A.,  Lyubarskii Y.~E.,  2000, A{\&}A, 355, 1168

\bibitem[\protect\citeauthoryear{Philippov \& Kramer}{Philippov \&
  Kramer}{2022}]{Philippov2022}
Philippov A.,  Kramer M.,  2022, \mn@doi [Ann. Rev. Astron. Astrophys.]
  {10.1146/annurev-astro-052920-112338}, 60, 495

\bibitem[\protect\citeauthoryear{Primak, Tiburzi, van Straten, Dyks  \&
  Gulyaev}{Primak et~al.}{2021}]{Primak2021}
Primak N.,  Tiburzi C.,  van Straten W.,  Dyks J.,   Gulyaev S.,  2021, A{\&}A,
  657, 34

\bibitem[\protect\citeauthoryear{Radhakrishnan \& Cooke}{Radhakrishnan \&
  Cooke}{1969}]{Radhakrishnan1969}
Radhakrishnan V.,  Cooke D.~J.,  1969, ApJ Lett., 3, 225

\bibitem[\protect\citeauthoryear{Ruderman \& Sutherland}{Ruderman \&
  Sutherland}{1975}]{Ruderman1975}
Ruderman M.~A.,  Sutherland P.,  1975, \mn@doi [ApJ] {10.1086/153393}, 196, 51

\bibitem[\protect\citeauthoryear{Weltevrede \& Johnston}{Weltevrede \&
  Johnston}{2008}]{Weltevrede2008}
Weltevrede P.,  Johnston S.,  2008, \mn@doi [MNRAS]
  {10.1111/j.1365-2966.2008.13950.x}, 391, 1210

\bibitem[\protect\citeauthoryear{von Hoensbroech, Lesch  \& Kunzl}{von
  Hoensbroech et~al.}{1998}]{VonHoensbroech1998}
von Hoensbroech A.,  Lesch H.,   Kunzl T.,  1998, A{\&}A, 336, 209

\makeatother
\end{thebibliography}

%%%%%%%%%%%%%%%%%%%%%%%%%%%%%%%%%%%%%%%%%%%%%%%%%%

%%%%%%%%%%%%%%%%% APPENDICES %%%%%%%%%%%%%%%%%%%%%

\appendix

\section{Mathematics of the partial-coherence model}
\label{app:maths}

\subsection{Mode notation}

We begin by mathematically defining the two orthogonal modes, and then consider their partially-coherent interaction. We begin by considering orthogonal modes that are fully linearly polarized and then generalize to the elliptical case in Section \ref{app:sec:ellip}. Without loss of generality we set up the basis describing the modes such that mode 1 is projected along the x-axis and mode 2 along the y-axis. In this set-up we can write the two modes as x-y column vectors as follows:
\begin{equation}
    M_{1} = A_{1}\sin{(\omega t + \phi_{1})}
    \begin{pmatrix}
    1 \\ 0
    \end{pmatrix}, \\
    M_{2} = A_{2}\sin{(\omega t + \phi_{2})}
    \begin{pmatrix}
    0 \\ 1
    \end{pmatrix}, 
\label{eq:modesintro}
\end{equation}
where mode amplitudes are $A_{1}$ and $A_{2}$, phase offsets are $\phi_{1}$ and $\phi_{2}$, $\omega$ is the wave frequency and $t$ represents time.

We are only interested in the relative mode amplitudes and phases which means we can drop the time varying term in subsequent equations. We can also define the mode amplitude ratio $\gamma = A_{2}/A_{1}$ and mode relative phase offset $\eta = \phi_{2}-\phi_{1}$. Without loss of generality we define the x-y basis such that the brighter of the two modes points along the x-axis. This means that $A_{2} \leq A_{1}$ so that $0 \leq \gamma \leq 1$. We re-write the mode equations in complex phase notation as 
\begin{equation}
    M_{1} = A_{1}\exp^{i\phi_{1}}
    \begin{pmatrix}
    1 \\ 0
    \end{pmatrix}, \\
    M_{2} = \gamma A_{1}\exp^{i(\phi_{1} + \eta)}
    \begin{pmatrix}
    0 \\ 1
    \end{pmatrix}.
\end{equation}
Again, as we are interested only in relative mode amplitudes and phases, we can choose to set $A_{1} = 1$ and $\phi_{1} = 0$, giving 
\begin{equation}
    M_{1} = 
    \begin{pmatrix}
    1 \\ 0
    \end{pmatrix}, \\
    M_{2} = \gamma\exp^{i\eta}
    \begin{pmatrix}
    0 \\ 1
    \end{pmatrix}.
\end{equation}

\subsection{Partially-coherent combination}

We define $C$ as the fraction of each mode that is added together coherently to produce a third mode. It is through this coherent addition that circular polarization is generated from the two linear modes. The remainder of each mode (a fraction $1-C$) remains as an incoherent contribution to the final observed polarization. For model simplicity, we assume that the same coherence fraction applies to both of the original orthogonal modes, rather than having a different parameter for each one. Applying this partially-coherent combination step generates three modes as follows:
\begin{equation}
    \begin{split}
        &M_{a} = (1-C)M_{1} \\
        &M_{b} = (1-C)M_{2} \\
        &M_{c} = C(M_{1} + M_{2}).
    \end{split}
\end{equation}
Writing these out in full, we obtain 
\begin{equation}
\begin{split}
    &M_{a} = 
    \begin{pmatrix}
    (1-C) + 0i \\
    0 + 0i
    \end{pmatrix} \\
    &M_{b} = 
    \begin{pmatrix}
    0 + 0i \\
    (1-C)\gamma(\cos{\eta} + i\sin{\eta})
    \end{pmatrix} \\
    &M_{c} = 
    \begin{pmatrix}
    C + 0i \\
    C\gamma(\cos{\eta} + i\sin{\eta})
    \end{pmatrix}
\end{split}
\end{equation}
as our three mode contributions to the observed polarimetric results. Thus our model is dependent on only three parameters: the mode amplitude ratio $\gamma$, mode phase offset $\eta$ and coherence fraction $C$. 

\subsection{Stokes parameters}

With the modes set up in this way, we can calculate the Stokes parameters associated with each individual mode contribution using the following equations:
\begin{equation}
    \begin{split}
        &I = \lvert E_{x}\rvert^{2} + \lvert E_{y}\rvert^{2} \\
        &Q = \lvert E_{x}\rvert^{2} - \lvert E_{y}\rvert^{2} \\
        &U = 2\Re{(E_{x}E_{y}*)} \\
        &V = -2\Im{(E_{x}E_{y}*)}.
    \end{split}
\end{equation}
This gives us
\begin{equation}
\begin{split}
    &I_{a} = (1-C)^{2} \\
    &Q_{a} = (1-C)^{2} \\
    &U_{a} = 0 \\
    &V_{a} = 0 \\
\end{split}
\end{equation}
\begin{equation}
\begin{split}
    &I_{b} = (1-C)^{2}\gamma^{2} \\
    &Q_{b} = -(1-C)^{2}\gamma^{2} \\
    &U_{b} = 0 \\
    &V_{b} = 0 \\
\end{split}
\end{equation}
\begin{equation}
\begin{split}
    &I_{c} = C^{2}(1 + \gamma^{2}) \\
    &Q_{c} = C^{2}(1 - \gamma^{2}) \\
    &U_{c} = 2\gamma C^{2}\cos{\eta} \\
    &V_{c} = 2\gamma C^{2}\sin{\eta} \\
\end{split}
\end{equation}

Summing together the three contributions gives the final Stokes parameters of the simulation as 
\begin{equation}
\begin{split}
    &I = I_{a} + I_{b} + I_{c} = \left((1-C)^{2} + C^{2}\right)\left(1 + \gamma^{2}\right) \\
    &Q = Q_{a} + Q_{b} + Q_{c} = \left((1-C)^{2} + C^{2}\right)\left(1 - \gamma^{2}\right) \\
    &U = U_{a} + U_{b} + U_{c} = 2\gamma C^{2}\cos{\eta} \\
    &V = V_{a} + V_{b} + V_{c} = 2\gamma C^{2}\sin{\eta} \\
\end{split}
\label{eq:Stokes}
\end{equation}

We have set up these simulated Stokes parameters only in terms of their relative amplitudes. They therefore do not give any information about the actual intensity of observed emission, making it impossible to compare them directly to the Stokes parameters measured for actual data. However, the relative relationships of the Stokes parameters are comparable: it is possible to compare Stokes parameters, or combinations thereof, provided we calculate them as fractions of total intensity.

When starting from a mathematical origin, as we have done here, it is logical to make use of the \textit{mode amplitude ratio}, $\gamma~=~A_{2}/A_{1}$, to describe the relationship between the amplitudes of the two electric fields of the modes. However, when considering the effects of the parameters, it brings greater clarity to consider the \textit{mode strength ratio}, $R~=~I_{b}/I_{a}$, in terms of a ratio of powers instead, as this is more directly comparable to the observable Stokes parameters. The relationship between the two is given by $R = \gamma^{2}$. 

\subsection{Generalization to elliptical modes and non-orthogonal modes}
\label{app:sec:ellip}

More generally, we can set up the two initial modes as being elliptical. Modifying equation \ref{eq:modesintro}:
\begin{equation}
    M_{1} = 
    \begin{pmatrix}
    A_{1, x}\sin{(\omega t + \phi_{1, x})} \\ A_{1, y}\sin{(\omega t + \phi_{1, y})}
    \end{pmatrix}, \\
    M_{2} = 
    \begin{pmatrix}
    A_{2, x}\sin{(\omega t + \phi_{2, x})} \\ A_{2, y}\sin{(\omega t + \phi_{2, y})}
    \end{pmatrix}, 
\end{equation}

As before, we set $\gamma = A_{2,y}/A_{1,x}$ and $\eta = \phi_{2,y}-\phi_{1,x}$, and we choose to set $A_{1,x} = 1$ and $\phi_{1,x} = 0$. We further define $\alpha = A_{1,y}/A_{1,x}$; $\beta = A_{2,y}/A_{2,x}$; $\chi = \phi_{1,y}$; and $\psi = \phi_{2,x} - \eta$. This gives:
\begin{equation}
    M_{1} = 
    \begin{pmatrix}
    1 \\ \alpha\exp^{i\chi}
    \end{pmatrix}, \\
    M_{2} = \gamma\exp^{i\eta}
    \begin{pmatrix} 
    \beta\exp^{i\psi} \\ 1
    \end{pmatrix}.
\end{equation}

The calculation of Stokes parameters from these more general modes then follows the same prescription as described above for the linear-mode case. The model Stokes parameters equivalent to those shown in equation \ref{eq:Stokes} are as follows:
\begin{align}
    \begin{split}
        I ={}& \left((1-C)^{2} + C^{2}\right)\left(1 + \gamma^{2} + \alpha^{2} + \beta^{2}\gamma^{2}\right) \\ 
            & + C^{2}\left(2\beta\gamma\cos{(\eta + \psi)} + 2\alpha\gamma\cos{(\eta - \chi)} \right) 
    \end{split} \\
    \begin{split}
        Q ={}& \left((1-C)^{2} + C^{2}\right)\left(1 - \gamma^{2} - \alpha^{2} + \beta^{2}\gamma^{2}\right) \\ 
            & + C^{2}\left(2\beta\gamma\cos{(\eta + \psi)} - 2\alpha\gamma\cos{(\eta - \chi)} \right) 
    \end{split} \\
    \begin{split}
        U ={}& 2 C^{2}\left(\gamma\cos{\eta} + \alpha\cos{\chi} + \alpha\beta\gamma\cos{(\eta + \psi - \chi)} + \beta\gamma^{2}\cos{\psi} \right) \\
            & + 2(1-C)^{2}\left(\alpha\cos{\chi} + \beta\gamma^{2}\cos\psi \right)
    \end{split} \\
    \begin{split}
        V ={}& 2 C^{2}\left(\gamma\sin{\eta} + \alpha\sin{\chi} - \alpha\beta\gamma\sin{(\eta + \psi - \chi)} - \beta\gamma^{2}\sin{\psi} \right) \\
            & + 2(1-C)^{2}\left(\alpha\sin{\chi} + \beta\gamma^{2}\sin\psi \right)
    \end{split} 
\label{eq:Stokesgeneralized}
\end{align}

We can see that we return to the linear mode case by setting $\alpha = \beta = 0$. There are two key additional cases to consider here: non-orthogonal modes and orthogonal elliptical modes. For the former, there is no relationship between the behaviours of modes 1 and 2, so the two new amplitude parameters $\alpha$ and $\beta$ can take any values between 0 and 1, and the two new phase parameters $\chi$ and $\psi$ can take any value between $0\degree$ and $360\degree$. A non-orthogonal mode set-up seems unlikely to be helpful, as the results it generates are completely general. To obtain orthogonal elliptical modes, we require $\alpha \equiv -\beta$ and $\chi \equiv -\psi$. Even in this case, the introduction of two additional parameters mean that the elliptical generalization is excessively flexible. We present this generalization for completeness, but the model is only really useful for comparison to observations and making predictions when restricted to the linear mode case.

\section{Calculation of $R$ in terms of fractional polarizations}
\label{app:appendixgamma}
We use equation~block~\ref{eq:Stokes} to write $l^{2} = (L_{d}/I_{d})^{2} = (L_{m}/I_{m})^{2} = (Q_{m}/I_{m})^{2}~+~(U_{m}/I_{m})^{2} = \left(\frac{1 - \gamma^{2}}{1 + \gamma^{2}}\right)^{2}~+~((V_{m}/I_{m})/\tan{\eta})^{2}$.
Using $\lvert v\rvert~=~\lvert V_{d}\rvert/I_{d}~=~\lvert V_{m}\rvert/I_{m}$ we can convert this to observational parameters: $\left(\frac{1 - \gamma^{2}}{1 + \gamma^{2}}\right)^{2} + ((V_{m}/I_{m})/\tan{\eta})^{2} = \left(\frac{1 - \gamma^{2}}{1 + \gamma^{2}}\right)^{2} + ((\lvert V_{d}\rvert/I_{d})/\tan{\eta})^{2} = \left(\frac{1 - \gamma^{2}}{1 + \gamma^{2}}\right)^{2} + (\lvert v\rvert/\tan{\eta})^{2}$. Rearranging this gives the mode strength ratio $R = \gamma^{2}$ in terms of observational parameters $l$ and $\lvert v\rvert$, and phase offset model parameter $\eta$, as
\begin{equation}
    R = \gamma^{2} = \frac{1-\sqrt{l^{2} - \left( \frac{\lvert v\rvert}{\tan\eta}\right)^{2}}}{1+\sqrt{l^{2} - \left( \frac{\lvert v\rvert}{\tan\eta}\right)^{2}}}.
    \label{eq:gamma}
\end{equation}
When $\eta = \uppi/2$, this simplifies to 
\begin{equation}
    R = \frac{1-l}{1+l}.
    \label{eq:gammasimple}
\end{equation}
This equation gives results in the range $0\leq\gamma\leq1$, as expected from the model set-up. It is valid for $\sqrt{l^{2} - \left( \frac{\lvert v\rvert}{\tan\eta}\right)^{2}}\leq 1$. The left side of that equation is maximised for $\eta = \uppi/2$, giving $0\leq l\leq 1$, which is already required from the physics. 

\section{Calculation of $C$ in terms of fractional polarizations}
\label{app:appendixC}
From equation~\ref{eq:Stokes}, we can write absolute circular polarization fraction $\lvert v\rvert = \lvert V_{d}\rvert/I_{d} = \lvert V_{m}\rvert/I_{m}$ as follows:
\begin{equation}
    \lvert v\rvert = \frac{\left\lvert2\gamma C^{2}\sin{\eta}\right\rvert}{\left((1-C)^{2} + C^{2}\right)\left(1 + \gamma^{2}\right)}.
\end{equation}
Now define $\alpha = 2\gamma\sin{\eta}/(1 + \gamma^{2})$ and substitute this in. From equation \ref{eq:gamma} we can see that $\alpha$ may also be written as $\alpha~=~\sqrt{(1-l^{2})\sin^{2}{\eta} + \lvert v\rvert^{2}\cos^{2}{\eta}}$, a convenient parameter for comparing to $\lvert v\rvert$. This gives 
\begin{equation}
    \lvert v\rvert = \frac{\lvert\alpha\rvert C^{2}}{\left((1-C)^{2} + C^{2}\right)}.
\end{equation}
Multiplying both sides by the right-hand side denominator and factoring out terms gives a quadratic equation: 
\begin{equation}
    (2\lvert v\rvert - \lvert\alpha\rvert)C^{2} + (-2\lvert v\rvert)C + \lvert v\rvert = 0,
    \label{eq:quadratic}
\end{equation}
which can be solved in the usual way with the quadratic formula. This gives the required result, but with two possible roots: 
\begin{equation}
    C_{\pm} = \frac{\lvert v\rvert \pm \sqrt{\lvert v\rvert(\lvert\alpha\rvert-\lvert v\rvert)}}{2\lvert v\rvert - \lvert\alpha\rvert}.
\end{equation}

For any pair of parameters $\lvert v\rvert$ and $l$ (defined such that $0\leq \lvert v\rvert\leq1$, $0\leq l\leq1$ and $l^{2} + \lvert v\rvert^{2} \leq 1$) we require the root for which $0\leq C\leq1$. We find that $C_{+} < 0$ for $\lvert v\rvert < \lvert \alpha\rvert/2$ and $C_{+} > 1$ for $\lvert v\rvert > \lvert \alpha\rvert/2$, whereas $0 \leq C_{-} \leq 1$ everywhere other than $\lvert v\rvert = \lvert \alpha\rvert/2$. Hence we quote $C_{-}$ in equation~\ref{eq:Csummary} for all cases other than other than at the divergent points $\lvert v\rvert = \lvert \alpha\rvert/2$. 

To solve for $C$ at these divergent points, return to equation~\ref{eq:quadratic} and substitute in $\lvert v\rvert = \lvert \alpha\rvert/2$. The term in $C^{2}$ vanishes and the resulting linear equation immediately gives $C = 1/2$. 

Note also that the square root in the numerator means that real answers are only possible for $\lvert \alpha\rvert \geq \lvert v\rvert$. This can be rearranged as $l^{2} + \lvert v\rvert^{2}(1 - \cos^{2}{\eta}) \leq 1$. The left hand side is largest when $\eta = \uppi/2$, giving $l^{2} + \lvert v\rvert^{2} \leq 1$, which is already a requirement from the physics.

% Don't change these lines
\bsp	% typesetting comment
\label{lastpage}
\end{document}